\documentclass{lmcs}
\pdfoutput=1

\usepackage{lastpage}
\lmcsdoi{17}{1}{18}
\lmcsheading{}{\pageref{LastPage}}{}{}%
{Jul.~10,~2017}{Mar.~02,~2021}{}

\usepackage[utf8]{inputenc}

\usepackage{amssymb}
\theoremstyle{definition}
\newtheorem{rema}[thm]{Remark}

\def\vspec#1{\special{ps:#1}}%  passes #1 verbatim to the output
\edef\zerocapsule#1{\vspec{gsave 1 1 scale scaleDvips #1 grestore}}

\newdimen\latticeUnit
\newdimen\boxMargin \newdimen\layerMargin

\newbox\tbox \newbox\ttbox
\chardef\tc=8 \chardef\tcc=9
\newdimen\tdimen \newdimen\ttdimen
\newdimen\sdimen \newdimen\ssdimen
\newtoks\ttoks \newtoks\tttoks
\newcount\tcount \newcount\ttcount
\newcount\tttcount \newcount\ttttcount

\newbox\diagbox
\newdimen\rightlim \newdimen\leftlim
\newdimen\upperlim \newdimen\lowerlim
\newdimen\dimA \newdimen\dimB
\newdimen\dimC \newdimen\dimD
\newdimen\dimE \newdimen\dimF \newdimen\dimG

\def\clist{\\}% `clist' is of the form {\\(2,3) (34.0pt,47.4pt)\\...}
\def\rightappend#1\to#2{\ttoks={#1\\}\tttoks=\expandafter{#2}%
 \edef#2{\the\tttoks\the\ttoks}}

{\catcode`p=12 \catcode`t=12 \gdef\gazonc#1pt{#1}}
\let\getfactor=\gazonc
\def\dimensionToNumber#1{\expandafter\getfactor\the#1}
\let\ptless=\dimensionToNumber

\def\nodebox#1{%
 \futurelet\com\nnodeboxx#1\with}

\def\nnodeboxx{%
 \ifx\com\invisible
  \let\next=\invisibleNodeBox
 \else
  \let\next=\visibleNodeBox
 \fi\next}

\def\visibleNodeBox#1\with#2#3{%
 \def\bm{#3}
 \setbox#2=\hbox{\kern\bm\vbox{\kern\bm\hbox{$\displaystyle{#1}$%
  }\kern\bm}\kern\bm}}

\def\invisible{}

\def\invisibleNodeBox#1\with#2#3{%
 \def\bm{#3}
 \setbox#2=\hbox{\kern\bm\vbox{\kern\bm\hbox{$\displaystyle{#1}$%
  }\kern\bm}\kern\bm}
 \setbox#2=\hbox{\vrule width0pt height\ht#2 depth\dp#2%
  \vbox to0pt{\hrule height0pt depth0pt width\wd#2}}}

\gdef\object(#1,#2)=#3{%
 \nodebox{#3}\tbox\boxMargin
 \tdimen=.5\wd\tbox \ttdimen=.5\ht\tbox \advance\ttdimen by.5\dp\tbox
 \edef\bbbox{(#1,#2)(\ptless{\tdimen}pt,\ptless{\ttdimen}pt)}%
 \expandafter\rightappend\bbbox\to\clist
 \advance\tdimen by-#1\latticeUnit \advance\ttdimen by-#2\latticeUnit
 \tdimen=-\tdimen \ttdimen=-\ttdimen
 \sdimen=\tdimen
  \ifdim\sdimen<\leftlim \global\leftlim=\sdimen \fi
  \advance\sdimen by\wd\tbox
  \ifdim\sdimen>\rightlim \global\rightlim=\sdimen \fi
 \sdimen=\ttdimen
  \ifdim\sdimen<\lowerlim \global\lowerlim=\sdimen \fi
  \advance\sdimen by\ht\tbox
  \ifdim\sdimen>\upperlim \global\upperlim=\sdimen \fi
 \put(\ptless{\tdimen},\ptless{\ttdimen}){\unhbox\tbox}
}

\def\refer(#1,#2)\to(#3,#4){\def\a{#1}\def\b{#2}%
 \tcount=#3 \ttcount=#4
 \expandafter\rreferr\clist\empty}
\def\rreferr\\#1{\ifx#1\empty \let\next=\relax
 \else \let\next=\rrreferrr \fi \next}
\def\rrreferrr#1,#2)(#3,#4){\def\aa{#1}\def\bb{#2}%
 \ifdim\a pt=\aa pt \ifdim\b pt=\bb pt
  \dimen\tcount=#3 \dimen\ttcount=#4
 \fi\fi \rreferr}

\gdef\edge{\let\lcommand=\line
 \morphismSwitch}

\gdef\morphism{\let\lcommand=\vector
 \morphismSwitch}

\def\morphismSwitch(#1,#2)to(#3,#4){%
 \def\mNext{\ifx\farg[\morphismBody(#1,#2)(#3,#4)%
  \else \slantMorphism(#1,#2)(#3,#4)\fi}
 \futurelet\farg\mNext}

\def\morphismBody(#1,#2)(#3,#4){%
 \slantMorphism(#1,#2)(#3,#4)%
 \expandafter\attachSwitch}

\def\abs#1{\ifdim#1<0pt-#1\else#1\fi}

\def\attachSwitch#1#2]{
 \nodebox{{\scriptstyle #2}}\tbox\boxMargin
 \rightsidefalse
 \diagonalfalse \horizontalfalse \verticalfalse
 \def\tNext{\ifx\targ[\expandafter\attachOption
  \else \attachBody\fi}
 \futurelet\targ\tNext}

\newif\ifrightside
\newif\ifhorizontal
\newif\ifvertical
\newif\ifdiagonal

\def\attachOption#1{%
 \attachOptionLoop}
\def\attachOptionLoop#1{%
 \ifx#1]
  \let\next\attachBody
 \else
  \ifx#1R \rightsidetrue \fi
  \ifx#1H \horizontaltrue \fi
  \ifx#1V \verticaltrue \fi
  \ifx#1D \diagonaltrue \fi
  \let\next\attachOptionLoop
 \fi
 \next
}

\newtoks\labelPosition
\labelPosition={.5}

\def\attachBody{%
 \dimC=\xC \advance\dimC by-\xB
 \dimD=\yC \advance\dimD by-\yB
 \dimA=\xB \advance\dimA by\the\labelPosition\dimC
 \dimB=\yB \advance\dimB by\the\labelPosition\dimD
 \ifhorizontal \ifdim0pt<\abs\dimD \attachBodyHorizontal \fi
 \else \ifvertical \ifdim0pt<\abs\dimC \attachBodyVertical \fi
 \else \ifdiagonal \attachBodyDiagonal
 \else
  \tdimen.2\dimD \tdimen=\abs\tdimen
  \ifdim\abs\dimC<\tdimen
   \attachBodyHorizontal
  \else \tdimen.2\dimC \tdimen=\abs\tdimen
  \ifdim\abs\dimD<\tdimen
   \attachBodyVertical
  \else
   \attachBodyDiagonal
  \fi\fi
 \fi\fi\fi}

\def\attachBodyHorizontal{%
 \dimG=.5\ht\tbox \advance\dimG by.5\dp\tbox
 \ifrightside \dimE=\dimD \else \dimE=-\dimD \fi
 \dimF=\dimC
 \ifdim\dimD<0.0pt
  \ifrightside \advance\dimA by-.5\wd\tbox
  \else \advance\dimA by.5\wd\tbox \fi
 \else
  \ifrightside \advance\dimA by.5\wd\tbox
  \else \advance\dimA by-.5\wd\tbox \fi
 \fi
 \displaceLabel(\ptless{\dimA},\ptless{\dimB})(\ptless{\dimE},\ptless{\dimF},\ptless{\dimG},1)
}

\def\attachBodyVertical{%
 \dimG=.5\wd\tbox
 \ifrightside \dimE=\dimC \else \dimE=-\dimC \fi
 \dimF=\dimD
 \ifdim\dimC<0.0pt
  \ifrightside \advance\dimB by.5\ht\tbox \advance\dimB by.5\dp\tbox
  \else \advance\dimB by-.5\ht\tbox \advance\dimB by-.5\dp\tbox \fi
 \else
  \ifrightside \advance\dimB by-.5\ht\tbox \advance\dimB by-.5\dp\tbox
  \else \advance\dimB by.5\ht\tbox \advance\dimB by.5\dp\tbox \fi
 \fi
 \displaceLabel(\ptless{\dimA},\ptless{\dimB})(\ptless{\dimE},\ptless{\dimF},\ptless{\dimG},-1)
}

\def\attachBodyDiagonal{%
 \ifdim\dimD<0.0pt
  \ifrightside
   \advance\dimA by-.5\wd\tbox \advance\dimA by.5\boxMargin
  \else
   \advance\dimA by.5\wd\tbox \advance\dimA by-.5\boxMargin
  \fi
 \else
  \ifrightside
   \advance\dimA by.5\wd\tbox \advance\dimA by-.5\boxMargin
  \else
   \advance\dimA by-.5\wd\tbox \advance\dimA by.5\boxMargin
  \fi
 \fi
 \ifdim\dimC<0.0pt
  \ifrightside
   \advance\dimB by.5\ht\tbox \advance\dimB by.5\dp\tbox
   \advance\dimB by-.5\boxMargin
  \else
   \advance\dimB by-.5\ht\tbox \advance\dimB by-.5\dp\tbox
   \advance\dimB by.5\boxMargin
  \fi
 \else
  \ifrightside
   \advance\dimB by-.5\ht\tbox \advance\dimB by-.5\dp\tbox
   \advance\dimB by.5\boxMargin
  \else
   \advance\dimB by.5\ht\tbox \advance\dimB by.5\dp\tbox
   \advance\dimB by-.5\boxMargin
  \fi
 \fi
 \displaceLabel(\ptless{\dimA},\ptless{\dimB})(1.0,0.0,0.0,0)
}

\def\displaceLabel(#1,#2)(#3,#4,#5,#6){%
 \put(#1,#2){%
  \vspec{gsave
   currentpoint currentpoint translate
   /adif {#5 #4 abs mul #3 div 8.3 mul} def
   /bdif {0.0} def
   #6 0 gt {adif bdif translate} {bdif adif translate} ifelse
   neg exch neg exch translate}%
  \hbox to0pt{\hss \vbox to0pt{\vss
   \hbox{\copy\tbox}%
   \vss}\hss}%
  \vspec{currentpoint grestore moveto}%
 }
}

\newdimen\hd \newdimen\vd \newdimen\cd \newdimen\md
\newdimen\xB \newdimen\yB
\newdimen\xC \newdimen\yC

\def\slantMorphism(#1,#2)(#3,#4){%
 \hd=#3\latticeUnit \advance\hd by-#1\latticeUnit
 \vd=#4\latticeUnit \advance\vd by-#2\latticeUnit
 \vectorPosition(#1,#2)(#3,#4)%
 \edef\arg{{\ptless{\xB}}{\ptless{\yB}}{\ptless{\xC}}{\ptless{\yC}}}
  \ifx\lcommand\vector
   \expandafter\drawLine\arg
   \expandafter\putTip\arg
  \else
   \expandafter\drawLine\arg
  \fi
}

\font\tenline = line10
\def\tip{\vbox to0pt{\hbox to0pt{{\tenline\char"36}\hss}\vss}}

\def\putTip#1#2#3#4{%
 \put(#3,#4){%
 \vspec{gsave
  currentpoint currentpoint translate
  /tanrot {50 10 sub 30 20 sub atan rotate} def
  #4 #2 eq not {#3 #1 sub #4 #2 sub atan rotate}
   {#1 #3 lt {90 rotate} {270 rotate} ifelse} ifelse
  neg exch neg exch translate}%
 \tip%
 \vspec{currentpoint grestore moveto}}}

\newif\iflayer
\layerfalse

\def\drawLine#1#2#3#4{%
 \iflayer
  \put(0,0){\zerocapsule{2.5 setlinewidth 1 setgray
   #1 #2 moveto #3 #4 lineto stroke
   .5 setlinewidth 0 setgray
   #1 #2 moveto #3 #4 lineto stroke}}%
 \else
  \put(0,0){\zerocapsule{#1 #2 moveto #3 #4 lineto stroke}}%
 \fi
}

{\catcode`p=12 \catcode`t=12 \gdef\hazonc#1.#2pt{#1}}
\def\toJnt#1{\expandafter\hazonc\the#1}
\let\getInt=\toJnt

\newdimen\xd \newdimen\yd \newdimen\xe \newdimen\ye

\def\vectorPosition(#1,#2)(#3,#4){%
 \ifdim \hd>0.0pt
  \ifdim \vd>0.0pt
   \obtainDelta(#1,#2)(#3,#4)
  \else
   \vd=-\vd
   \obtainDelta(#1,#2)(#3,#4)
   \vd=-\vd \yd=-\yd \ye=-\ye
  \fi
 \else
  \hd=-\hd
  \ifdim \vd>0.0pt
   \obtainDelta(#1,#2)(#3,#4)
   \xd=-\xd \xe=-\xe
  \else
   \vd=-\vd
   \obtainDelta(#1,#2)(#3,#4)
   \vd=-\vd \xd=-\xd \xe=-\xe \yd=-\yd \ye=-\ye
  \fi
  \hd=-\hd
 \fi
 \xB=#1\latticeUnit \advance\xB by\xd
 \yB=#2\latticeUnit \advance\yB by\yd
 \xC=#3\latticeUnit \advance\xC by\xe
 \yC=#4\latticeUnit \advance\yC by\ye
}

\def\obtainDelta(#1,#2)(#3,#4){%
 \refer(#1,#2)\to(\tc,\tcc)%
 \ifdim\hd<\vd % if vd is greater, try to divide with it first
  \dimA=\dimen\tcc
  \multiply\dimA by\getInt\hd
  \divide\dimA by\getInt\vd    % dimA = tcc * hd / vd
  \dimB=\dimen\tc
  \ifdim \dimB<\dimA
   \dimA=\dimen\tc
   \multiply\dimA by\getInt\vd
   \divide\dimA by\getInt\hd   % dimA = tcc * vd / hd
   \xd=\dimen\tc \yd=\dimA
  \else
   \xd=\dimA \yd=\dimen\tcc
  \fi
 \else % if hd is greater
  \dimA=\dimen\tc
  \multiply\dimA by\getInt\vd
  \divide\dimA by\getInt\hd    % dimA = tcc * vd / hd
  \dimB=\dimen\tcc
  \ifdim \dimB<\dimA
   \dimA=\dimen\tcc
   \multiply\dimA by\getInt\hd
   \divide\dimA by\getInt\vd   % dimA = tcc * hd / vd
   \xd=\dimA \yd=\dimen\tcc
  \else
   \xd=\dimen\tc \yd=\dimA
  \fi
 \fi
 \refer(#3,#4)\to(\tc,\tcc)%
 \ifdim\hd<\vd % if vd is greater, try to divide with it first
  \dimA=\dimen\tcc
  \multiply\dimA by\getInt\hd
  \divide\dimA by\getInt\vd    % dimA = tcc * hd / vd
  \dimB=\dimen\tc
  \ifdim \dimB<\dimA
   \dimA=\dimen\tc
   \multiply\dimA by\getInt\vd
   \divide\dimA by\getInt\hd   % dimA = tcc * vd / hd
   \xe=-\dimen\tc \ye=-\dimA
  \else
   \xe=-\dimA \ye=-\dimen\tcc
  \fi
 \else % if hd is greater
  \dimA=\dimen\tc
  \multiply\dimA by\getInt\vd
  \divide\dimA by\getInt\hd    % dimA = tcc * vd / hd
  \dimB=\dimen\tcc
  \ifdim \dimB<\dimA
   \dimA=\dimen\tcc
   \multiply\dimA by\getInt\hd
   \divide\dimA by\getInt\vd   % dimA = tcc * hd / vd
   \xe=-\dimA \ye=-\dimen\tcc
  \else
   \xe=-\dimen\tc \ye=-\dimA
  \fi
 \fi
}

\def\ptToCoord#1{%
 \tcount=\expandafter\toInt\number\ptless\latticeUnit
 \divide#1 by\tcount}

\newenvironment{diagramme}{%
  \latticeUnit=100pt \boxMargin=3pt \layerMargin=3pt
  \rightlim=0pt \leftlim=0pt \upperlim=0pt \lowerlim=0pt
  \setbox\diagbox=\hbox\bgroup
  \begin{picture}(0,0)}%
 {\end{picture}\egroup
  \tdimen=\rightlim \advance\tdimen by-\leftlim
  \ttdimen=\upperlim \advance\ttdimen by-\lowerlim
  \begin{picture}(\ptless{\tdimen},\ptless{\ttdimen})%
                 (\ptless{\leftlim},\ptless{\lowerlim})
   \put(0,0){\unhbox\diagbox}
  \end{picture}%
}

\def\mor#1{\mathrel{\mathop{\longrightarrow}\limits^{\vbox
 to0pt{\vss \hbox{$\scriptstyle #1$}\kern-2pt}}}}
\def\leftmor#1{\mathrel{\mathop{\longleftarrow}\limits^{\vbox
 to0pt{\vss \hbox{$\scriptstyle #1$}\kern-2pt}}}}
\def\twomor#1{\mathrel{\mathop{\Longrightarrow}\limits^{\vbox
 to0pt{\vss \hbox{$\scriptstyle #1$}\kern-1pt}}}}
\def\shortmor#1{\mathrel{\mathop{\rightarrow}\limits^{\vbox
 to0pt{\vss \hbox{$\scriptstyle #1$}\kern-2pt}}}}
\def\shortleftmor#1{\mathrel{\mathop{\leftarrow}\limits^{\vbox
 to0pt{\vss \hbox{$\scriptstyle #1$}\kern-2pt}}}}
\def\fitarr#1{\vcenter{\offinterlineskip
 \halign{\hfil ##\hfil &$##$\hfil \crcr
 \vbox to0pt{\vss \hbox{\kern2pt$\scriptstyle #1$\kern3pt}}\crcr
 \noalign{\vskip1pt}
 \leaders\hrule height.7pt depth-.3pt\hfill $\mkern2mu$
  &\vbox to0pt{\vss \llap{$\rightarrow$}\vss} \crcr}}}

\def\sqda#1#2#3#4#5#6#7#8{%
 \let\ss=\scriptstyle
 \latticeUnit=1pt \boxMargin=3pt
 \object(0,\vert)={#1}
 \object(\hori,\vert)={#2}
 \object(0,0)={#3}
 \object(\hori,0)={#4}
 \morphism(0,\vert)to(\hori,\vert)[\ss{#5}]
 \morphism(0,\vert)to(0,0)[\ss{#6}][R]
 \morphism(\hori,\vert)to(\hori,0)[\ss{#7}]
 \morphism(0,0)to(\hori,0)[\ss{#8}][R]
}

\def\llapem#1#2{\llap{\hbox to#1em{\rm #2\hss}}}
\let\tempar\par \def\par{{\tempar}}%

\def\uplet#1#2#3{#1\mathbin\upharpoonright #2\mathbin{:=}#3}%

\def\lc{(\mskip-4mu|}%
\def\rc{|\mskip-4mu)}%

\newbox\tcd\setbox\tcd=\hbox{$\omega$}
\newcommand{\agemo}{\rotatebox[origin=c]{180}{$\omega$}}

\def\dbot{\bot\,\llap{$\bot$}}
\def\sdbot{\bot\,\llap{$\scriptstyle\bot$}}

\def\setplace#1{%
 \hbox to0pt{\scriptsize\hss
  \vbox{\offinterlineskip\halign{\hfil $##$\hfil\crcr
   \uparrow\crcr \noalign{\vskip2pt}#1\crcr}}\hss}}

\def\lbra#1{%
 \vtop{\hbox{\strut}%
  \kern-\baselineskip\kern-\lineskip
  \hbox{$\left\{\vcenter
   {\halign{\strut $\>##$\qquad \hfil &##\hfil \crcr
 #1}}\right.$}}}

\def\lg{\{\mskip-4.5mu[}%
\def\rg{]\mskip-4.5mu\}}%

\begin{document}

\title[Complete Call-by-Value Calculi of Control Operators, II]%
 {Complete Call-by-Value Calculi of Control Operators,~II: Strong termination}

\author[Ryu Hasegawa]{Ryu Hasegawa}	%required
\address{Graduate School
 of Mathematical Sciences, The University of Tokyo, Komaba 3-8-1,
 Meguro-ku, Tokyo 153-8914, Japan}	%required
\email{ryu@ms.u-tokyo.ac.jp}  %optional
%\thanks{thanks 1, optional.}	%optional

%% required for running head on odd and even pages, use suitable
%% abbreviations in case of long titles and many authors:

%% mandatory lists of keywords and classifications:
%\titlecomment{OPTIONAL comment concerning the title, \eg, if a variant
%or an extended abstract of the paper has appeared elsewehere}
%%%%%%%%%%%%%%%%%%%%%%%%%%%%%%%%%%%%%%%%%%%%%%%%%%%%%%%%%%%%%%%%%%%%%%%%%%%

%% the abstract has to PRECEED the command \maketitle:
%% be sure not to issue the \maketitle command twice!

\begin{abstract}
  \noindent
We provide characterizations of the strong termination property
 of the CCV (complete call-by-value) $\lambda\mu$-calculus introduced in
 the first part of this series of papers.
The calculus is complete with respect to the standard
 continuation-passing style (CPS) semantics.
The union-intersection type system for the calculus was developed
 in the first paper.
We characterize the strong normalizability of the calculus
 in terms of the CPS semantics and typeability.
\end{abstract}

\maketitle

%% start the paper here:
\section*{Introduction}

This is the second half of a series of papers.
In the first part, we proposed a
 call-by-value $\lambda\mu$-calculus, called
 the CCV $\lambda\mu$-calculus,
 which is complete for the
 continuation-passing style (CPS) semantics \cite{tsd43}.
Furthermore, we proposed the union-intersection type discipline.
Among others, we verified the following:
\begingroup
	\vskip0ex
        \hangafter0\hangindent1.8em
        \noindent
\llapem{1.8}{(1)}%
A term $M$ terminates with respect
 to the call-by-value evaluation if and only if its CPS translation
 $[\![M]\!]$ is solvable.

	\vskip0ex
        \noindent
\llapem{1.8}{(2)}%
A term $M$ is weakly normalizing if and only if its CPS translation $[\![M]\!]$
 is weakly normalizing.

	\vskip0ex
        \noindent
\llapem{1.8}{(3)}%
A term $M$ terminates with respect
 to the call-by-value evaluation if and only if $M$ is
 typeable.

	\vskip0ex
        \noindent
\llapem{1.8}{(4)}%
A term $M$ is weakly normalizing if and only if $M$ is typeable where
 the typing judgment of $M$ contains
 neither empty intersection nor empty union.

	\vskip.5ex
\endgroup
\noindent
The theme of the second part is to further exntend
 these results.
We give characterizations of the strong termination property.
Specifically, we show that

	\vskip.5ex
\begingroup
	\vskip0ex
        \hangafter0\hangindent1.8em
        \noindent
\llapem{1.8}{(5)}%
$M$ is strongly normalizing if and only if its CPS translation $[\![M]\!]$ is
 strongly normalizing.

\noindent
\llapem{1.8}{(6)}%
$M$ is strongly normalizing if and only if $M$ is typeable using
 empty intersection or empty union nowhere.

	\vskip0ex
\endgroup
\noindent
After a brief introduction to the CCV $\lambda\mu$-calculus and its type
 system containing union and intersection,
 we first verify the strong termination of
 a fragment in \S\ref{cud34} as an intermediate step.
Employing this result,
 we verify (6) in Thm.~\ref{rzq50} and (5) in Thm.~\ref{tml91}
 in \S\ref{eqw96}.

An essential idea behind our calculus is a
 departure from the conventional demand that
 terms be freely generated by syntactic grammars.
An analogy is found in arithmetic.
We frequently use expressions such as $3+5+7$.
When we calculate it (or implement a calculator on
 a computer), we forcibly interpret
 the expression as either $(3+5)+7$ or $3+(5+7)$ and
 perform the calculation.
However, strictly distinguishing between the two cases presents
 little advantage for humans to
 understand the essence of arithmetic.
We positively use the expression $3+5+7$ as an amalgamation
 of two types of bracketed expressions, or even as a sum of three numbers.
This type of intended ambiguity helps us process arithmetic flexibly.
In the same vein, we introduce ambiguity in the constructors of
 call-by-value calculus.
The resulting calculus is complete
 with respect to the standard semantics, and yet usable.

\section{Preliminaries}\label{muj01}

We recall the CCV $\lambda\mu$-calculus
 and the union-intersection type discipline for the calculus \cite{tsd43}.
We also review some of the results needed later.
Details are found in the first paper.

\subsection{CCV $\lambda\mu$-calculus}\label{yhd30}

The CCV $\lambda\mu$-calculus is a variant of
 the call-by-value $\lambda\mu$-calculus.
It employs the let-syntax as in Moggi's $\lambda_c$-calculus \cite{dvp96}.
However, we write the let-binding to the right of its body:
\[
\uplet MxN\qquad\text{in place of}\qquad{\sf let}\ x=N\ {\sf in}\ M.
\]
Note that the order of $M$ and $N$ is reversed.

Formally, the syntax of the CCV $\lambda\mu$-calculus
 is given as follows.
We distinguish ordinary variables and continuation variables.
We also distinguish terms $M$ and jumps $J$.
They are defined mutually recursively by the following syntax:
\begin{align*}
    M \mathrel{::=}&~ x \ \mid\  \lambda x.\,M \ \mid\  MM \ \mid\  \uplet MxM \ \mid\  \mu k.\,J \\
  J \mathrel{::=}&~ [k]M \ \mid\  \uplet JxM
\end{align*}
where $x$ ranges over ordinary variables and $k$ over continuation
variables.
The notion of free variables is naturally defined.
The let-construct $\uplet MxN$ binds $x$, and its scope is $M$.
We use the notation $z\in M$ to denote that a variable $z$ that is
 ordinary or continuation occurs freely in $M$.

A key idea is that this syntax is not regarded
 to freely generate the entities.
We introduce syntactic equalities by two associativity axioms:
\begin{align*}
    \uplet Lx{(\uplet MyN)} =&~ \uplet{(\uplet LxM)}yN &\text{if } y\not\in L \\
    [k](\uplet LxM) =&~ \uplet{([k]L)}xM
\end{align*}
 where $L,\,M$, and $N$ are terms.
We do not syntactically distinguish two terms (or two jumps) if they turn
 out to be equal by a series of application of these rules.
We mostly omit brackets:
\begin{gather*}
 \uplet Lx{\uplet MyN}\\
 \uplet{\mu k.\,J}xM\\
 [k]\uplet LxM.
\end{gather*}
For the first, if the side condition is not satisfied (i.e.,
 if $y\in L$), we take it to mean
 $\uplet{(\uplet LxM)}yN$.
For the second, we read it as $\mu k.\,(\uplet JxM)$.

A {\it value} is either a variable or a lambda abstraction.
We have the following ten reduction rules, where $N$ is a non-value
 and $V$ is a value:
 \[
\halign{\hspace{6em}\rlap{\footnotesize $(#)$}&\kern5em
  $#$\hfil &${}\ \rightarrow\ #$\kern2em\hfil &#\hfil \cr
 {\it ad}_1 & NM & \uplet{zM}zN \cr
 {\it ad}_2 & VN & \uplet{Vz}zN \cr
 \beta_\lambda & (\lambda x.\,M)V & \uplet MxV \cr
 \beta_{\it let} & \uplet MxV & M\{V/x\} \cr
 \beta_\mu & \uplet Mx\mu k.\,J & \rlap{$\mu k.\,J\{[k]\square\mapsto
  [k]\uplet Mx\square\}$}\cr
 \beta_{\it jmp} & [l]\mu k.\,J & J\{l/k\}\cr
 \eta_\lambda & \lambda x.\,Vx & V &(if $x\not\in V$)\cr
 \eta_{\it let} & \uplet xxM & M\cr
 \eta_\mu & \mu k.\,[k]M & M &(if $k\not\in M$)\cr
 {\it exch} & \uplet{(\mu k.\,J)}xM & \mu k.\,\uplet JxM &(if $k\not\in M$)\cr
}\]
In the first two rules,
 $z$ is a fresh ordinary variable.
We use braces for substitution.
The notation $J\{[k]\square\mapsto [k]\uplet Mx\square\}$ is
 a standard context substitution in the $\lambda\mu$-calculus.
We write $L=_{\it ccv}M$ if two terms are equivalent regarding
 the smallest equivalence relation generated from reductions.

Since we are allowed to alter the scope of let-binding by the equality axioms,
 the continuation $\uplet Mx\square$ captured by rule $\beta_\mu$
 is changeable, depending on the choice of the scope.
This ambiguity is intended and is crucial to verifying the
 sharpened completeness theorem, which
 is a key result in our previous paper.

\begin{rema}\label{nrk88}
In the previous paper, we had the third equality axiom:
\[
\uplet{(\mu k.\,J)}xM\ =\ \mu k.\,(\uplet JxM)\qquad\text{ if } k\not \in M.
\]
Namely, we were able to exchange the $\mu$-operator and let-operator.
As we commented in the first paper,
 however, this equality axiom was inessential for proving
 the sharpened completeness theorem.
Instead, we can consider a reduction rule.
For the purpose of this paper, the latter approach gives
 better results at this stage.
Therefore, we adopt the rule {\it exch} above.
It is open whether the same results are obtained if the equality axiom
 is chosen.
See Rem.~\ref{orv02}.
\end{rema}

The semantics of the CCV $\lambda\mu$-calculus is given by the
 call-by-value CPS translation.
It maps each CCV term and each jump to a lambda term.
In place of a standard CPS translation,
 we adopt the definition via the colon translation,
 which is introduced to diminish a number of
 superfluous redexes \cite{pag41}.
The latter behaves better in regard to reductions, which are our
 concern.
We use the notation $\lc M\rc[K]$ in place of $M:K$ for
 readability.
\[
\halign{\kern5em $#$\hfil &${}\ \mathrel{:=}\ #$\hfil\cr
 \lc V\rc[K] & KV^*\cr
 \lc V_1V_2\rc[K] & V_1^*V_2^*K\cr
 \lc VN\rc [K] & \lc N\rc [\lambda y.\,V^*yK]\cr
 \lc NV\rc [K] & \lc N\rc [\lambda x.\,xV^*K]\cr
 \lc N_1N_2\rc [K] & \lc N_1\rc [\lambda x.\,\lc N_2\rc [
  \lambda y.\,xyK]]\cr
 \lc \uplet LxM\rc [K] & \lc M\rc [\lambda x.\,\lc L\rc [K]]\cr
 \lc \mu k.\,J\rc [K] & (\lambda k.\,\lc J\rc )K\cr
	\noalign{\vskip1ex}
 \lc [k]M\rc  & \lc M\rc [k]\cr
 \lc \uplet JxM\rc  & \lc M\rc [\lambda x.\,\lc J\rc ]\cr
 \noalign{\vskip2ex}
 x^* & x\cr
 (\lambda x.\,M)^* & \lambda xk.\,\lc M\rc [k]\cr
 \noalign{\vskip2ex}
 [\![M]\!] & \lambda k.\,\lc M\rc [k]\cr
}
\]
Here $V$ and $V_i$ are values, while $N$ and $N_i$ are non-values.
$L$ and $M$ are terms.
The results are applicable to the
 standard CPS translation, with some extra arguments, as discussed in
 Rem.~\ref{zqb05}.

It is better to regard the target of the CPS translation
 as a sorted lambda calculus.
There are four sorts.
The terms of the sorted calculus are defined as follows:
\[
\halign{\kern5em #\kern3em\hfil &$#$\hfil &${}\ \mathrel{::=}\ #$\hfil\cr
 Term& T & \lambda k.\,Q\ \ |\ \ WW\cr
 Jump& Q & KW\ \ |\ \ TK\cr
 Value& W & x\ \ |\ \ \lambda x.\,T\cr
 Continuation& K & k\ \ |\ \ \lambda x.\,Q\cr}
\]
The CPS translation yields terms that are subject to this syntax.
$\lc M\rc[K]$ and $\lc J\rc$ produce terms of sort $Q$, and
 $V^*$ of sort $W$.
$[\![M]\!]$ has sort $T$.

We also have the inverse translation $(\hbox{-})^{-1}$ from
 the target calculus back into the CCV $\lambda\mu$-calculus.
We do not, however, need the concrete shape of the translation,
 for we use it only through Lem.~\ref{bdf72} and \ref{jzz08} below.
We refer the interested reader to \cite{tsd43}.

We list several results that are needed in this paper from \cite{tsd43}.
We call a reduction by rule $(\eta_\mu)$ {\it vertical},
 and a reduction by $({\it ad}_1)$ or $({\it ad}_2)$
 {\it administrative}.
A non-administrative reduction is called {\it practical}.

\begin{lem}\label{bdf72}
Let $M$ be a term of the CCV $\lambda\mu$-calculus.
There is a term $M^\dagger$ of the calculus such that
 $M\mor{A^*}M^\dagger\leftmor{V^*}[\![M]\!]^{-1}$ (notice the direction).
Here $A^*$ denotes a finite number of administrative reductions
 and $V^*$ a finite number of vertical reductions.
\end{lem}

\begin{lem}\label{jzz08}
If $P\rightarrow_{\beta\eta}Q$ in the target calculus, then $P^{-1}\mor+Q^{-1}$
 by one or more steps of practical reductions with no use of rule {\it exch}.
\end{lem}

\begin{lem}\label{qck48}
If $L'\leftmor{V^*}L\mor{P^*}M$ by finite steps
 of vertical reductions and practical
 reductions, there is a term $M'$ such that $L'\mor{P^*}M'\leftmor{V^*}M$.
Moreover, the arrow $L'\mor{P^*}M'$ may be an identity only
 when $L\mor{P^*}M$ consists solely of $\beta_\mu$, $\beta_{\it jmp}$, $\eta_\mu$,
 and {\it exch}.
\end{lem}

In the verifications of Lem.~\ref{bdf72} and \ref{jzz08},
 we need the equality axioms of our calculus.
Sharpened completeness
 is an immediate consequence of these lemmata.
This explains why we include the equality axioms
 in the CCV $\lambda\mu$-calculus.
These lemmata are also necessary to prove one direction of implications
 in the main theorems.

\subsection{Union-intersection type discipline}

Types are divided into three categories:
 raw types $R$, subsidiary types $S$, and types $T$.
These are defined by the following syntax:
\begin{align*}
  R \mathrel{::=}&~ \alpha\mid S\rightarrow T \\
  S \mathrel{::=}&~ \textstyle{\bigcap}\, R \\
  T \mathrel{::=}&~ \textstyle{\bigcup}\, S
\end{align*}
where $\alpha$ ranges over atomic types.
$\bigcap R$ means a nonempty finite formal intersection $R_1\cap R_2\cap
\cdots\cap R_n$ ($n\geq 1$).
$\bigcup S$ is similar.
Intersection and union follow associativity and commutativity.

\begin{rema}\label{ibl41}
In our previous work \cite{tsd43}, the empty intersection $\omega$ and
 the empty union $\agemo$ are allowed.
In this paper, we use only the type derivations that
 contain $\omega$ or $\agemo$ nowhere.
So we omit them from the beginning.
\end{rema}

We define the subtype relation $\leq$ by the following derivation rules:
\begin{gather*}
\vbox{\offinterlineskip
 \halign{\strut $#$\hfil\cr
  \noalign{\hrule}
  \alpha\leq \alpha\cr
}}\kern3em
\vbox{\offinterlineskip
 \halign{\strut $#$\hfil\cr
  S'\leq S\qquad T\leq T'\cr
  \noalign{\hrule}
  S\rightarrow T\leq S'\rightarrow T'\cr
}}
\\
\vbox{\offinterlineskip
 \halign{\strut $#$\hfil\cr
  S\leq S'\cr
  \noalign{\hrule}
  S\cap S''\leq S'\cr
}}\kern3em
\vbox{\offinterlineskip
 \halign{\strut $#$\hfil\cr
  [\ S\leq S_i\ ]_i\cr
  \noalign{\hrule}
  S\leq \bigcap_iS_i\cr
}}
\\
\vbox{\offinterlineskip
 \halign{\strut $#$\hfil\cr
  T\leq T'\cr
  \noalign{\hrule}
  T\leq T''\cup T'\cr
}}\kern3em
\vbox{\offinterlineskip
 \halign{\strut $#$\hfil\cr
  [\ T_i\leq T\ ]_i\cr
  \noalign{\hrule}
  \bigcup_iT_i\leq T\cr
}}.
\end{gather*}
The notation $[\>S\leq S_i\>]_i$ means
 a sequence of derivations where $i$ ranges over a finite index set.
$[\>T_i\leq T\>]_i$ is similar.

A {\it typing judgment} has the form $\Gamma\;\vdash\;M\mathbin:T\;\mathbin|
 \;\Delta$ where $\Gamma$ is
 a finite sequence of $x_i\mathbin:S_i$, and $\Delta$
 is a finite sequence of $k_j\mathbin:T_j$.
Note that ordinary variables have only subsidiary types.
We assume a special type $\dbot$ for typing jumps.
The inference rules are given as follows:
\begin{gather*}
  \frac{}{\Gamma,x\mathbin:S\vdash x\mathbin:S \mid \Delta} \\[.5em]
\frac{[\Gamma,x\mathbin:S_i\vdash M\mathbin:T_i \mid \Delta ]_{i}}
{\Gamma \vdash \lambda x.\,M\mathbin:\bigcap_{i}(S_i\rightarrow T_i) \mid \Delta}
\\[.5em]
\frac{\Gamma \vdash M\mathbin:\bigcup_{i}\bigcap_{j}(S_{ij}\rightarrow T)
  \mid \Delta\qquad [ \Gamma \vdash N\mathbin:\bigcup_{j}S_{ij} \mid \Delta ]_{i}}{\Gamma \vdash MN\mathbin:T \mid \Delta}
\\[.5em]
\frac{[\Gamma,x\mathbin:S_i \vdash M\mathbin:T
  \mid \Delta ]_{i}\qquad \Gamma \vdash N\mathbin:\bigcup_{i}S_i \mid
\Delta}{\Gamma \vdash \uplet MxN\mathbin:T \mid \Delta}
\\[.5em]
\frac{\Gamma \vdash J\mathbin:\dbot \mid \Delta,k\mathbin:T}
{\Gamma \vdash \mu k.\,J\mathbin:T \mid \Delta}
\\[.5em]
\frac{\Gamma \vdash M\mathbin:T
  \mid \Delta,k\mathbin:T}{\Gamma \vdash [k]M\mathbin:\dbot \mid \Delta, k\mathbin:T}
\\[.5em]
\frac{[\Gamma,x\mathbin:S_i \vdash J\mathbin:\dbot
   \mid \Delta ]_{i}\qquad
  \Gamma \vdash N\mathbin:\bigcup_{i}S_i \mid \Delta}{\Gamma \vdash \uplet JxN\mathbin:\dbot \mid \Delta}
\\[.5em]
\frac{\Gamma \vdash M\mathbin:T \mid \Delta\qquad T\leq T'}{\Gamma \vdash M\mathbin:T' \mid \Delta}
\end{gather*}
Each index ($i$ or $j$) ranges over a finite set.
The notation $[\cdots]_i$ denotes a finite sequence of judgments.
In each rule, the same indices are understood to range over the same set.
For example, in the third rule, $i$ ranges over a finite set $I$ and
 $j$ over a finite set $J(i)$ depending on $i$, and these $I$ and $J(i)$ are
 shared between the assumptions.

\subsection{Type system of the target calculus}\label{ehd63}
The characterization of strong termination is verified through
 the type theory of the target calculus we present here.
The type theory is based on the standard intersection type discipline.
We assume a special atomic type $\dbot$ and write
 $\neg(\hbox{-})$ in place of $(\hbox{-})\rightarrow\dbot$.
We define the strict types $\tau,\kappa$, and $\sigma$ and types
 $\underline\kappa$ and $\underline\sigma$ by the following:
\begin{align*}
  \sigma \mathrel{::=}&~ \alpha\ \mid\, \underline\sigma\rightarrow\tau & \underline\sigma \mathrel{::=}&~ \textstyle{\bigcap}\,\sigma\\
  \kappa \mathrel{::=}&~ \neg\underline\sigma & \underline\kappa \mathrel{::=}&~ \textstyle{\bigcap}\,\kappa\\
  \tau \mathrel{::=}&~ \neg\underline\kappa
\end{align*}
where $\alpha$ represents atomic types.
$\bigcap\kappa$ denotes a finite formal intersection
 $\kappa_1\cap\kappa_2\cap\cdots\cap\kappa_n$ with $n\geq 1$.
$\bigcap\sigma$ is similar.
Intersection follows associativity and commutativity.
We have the subtype relation $\leq$ between types,
 which is defined naturally.

A typing judgment $\Pi,\Theta\vdash_s M\mathbin:\rho$ has
 two environments, $\Pi$ and $\Theta$,
 the former a finite sequence of $x\mathbin:\underline\sigma$
 and the latter of $k\mathbin:\underline\kappa$.
Here $M$ is either of term $T$, jump $Q$, value $W$, and continuation $K$,
 on which the kind of type $\rho$ depends.
A term has type $\tau$, a jump $\dbot$, a value $\sigma$, and a continuation $\kappa$.
We note that only strict types occur on the right
 hand of $\vdash_s$ (this is indicated by the subscript).

The inference rules of the intersection type discipline are
 standard \cite{bkw61}, except that the sorts must be respected.
For instance, the derivation rule for $W_1W_2$ is
\[
  \frac{\Pi,\Theta \vdash_s W_1\mathbin:\underline\sigma\rightarrow\tau\qquad
    [\Pi,\Theta \vdash_s W_2\mathbin:\sigma_i ]_i}{\Pi,\Theta \vdash_s W_1W_2\mathbin:\tau}
\]
Moreover, we include the inheritance rule for each sort; e.g.,
\[
\frac{\Pi,\Theta \vdash_s T\mathbin:\tau\qquad \tau\leq\tau'}{\Pi,\Theta \vdash_s T\mathbin:\tau'}
\]
to deal with $\eta$-rules.
We refer the reader to \cite{tsd43} for the presentation of the complete set.

\begin{thm}\label{cnp96}
Let $M$ be a term of the CCV $\lambda\mu$-calculus.
If $\Pi,\,\Theta\>\vdash_s\>[\![M]\!]\mathbin:\tau$ is derivable in the target calculus,
 then $\Gamma\>\vdash\>M\mathbin:T\>|\>\Delta$ is derivable 
 for some $\Gamma,\Delta$, and $T$ in the CCV $\lambda\mu$-calculus.
\end{thm}

\begin{proof}
The theorem is verified in \cite{tsd43} using the inverse translation.
By inspection of the proof therein, we see that
 the induced derivation of $\Gamma\>\vdash\>M\mathbin:T\>|\>\Delta$
 contains neither $\omega$ nor $\agemo$, provided the
 derivation of $\Pi,\,\Theta\>\vdash_s\>[\![M]\!]\mathbin:\tau$ contains
 $\omega$ nowhere.
\end{proof}

\section{Strong Normalization}\label{vis62}

The main theorems of this section are the following:
(i) a CCV $\lambda\mu$-term $M$ is strongly normalizable
 if and only if $[\![M]\!]$ is strongly normalizable (Thm.~\ref{tml91}), and
(ii) $M$ is strongly normalizable if and only if
 $M$ is typeable (Thm.~\ref{rzq50}).
We note that the strong normalizability of lambda terms is not
 closed under $\beta\eta$-equality.
It happens that $M=_{\beta\eta}N$ and $M$ is strongly normalizable,
 whereas $N$ is not.
Hence the main result (i) is sensitive to the choice of the
 CPS translation.
The following argument assumes
 the colon translation as defined in Preliminaries.
The case of the standard CPS translation is briefly discussed
 in~\ref{zqb05}.

\subsection{Termination of $\beta_\mu\beta_{jmp}\eta_\mu$
 reduction sequences}\label{cud34}
We verify the termination of all reduction sequences
 consisting exclusively of $\beta_\mu$, $\beta_{\it jmp}$, and $\eta_\mu$.
The reasons we prove it independently are the following:
(a) it is used later in Prop.~\ref{ntw00},
(b) it holds regardless of types,
(c) in the proof, we introduce the notion of places, which
 are used in the sequel, and
(d) it explains why we regard the exchange of
 the $\mu$-operator and let-operator
 as a reduction rule rather than an equality rule in our previous
 work.

Since we have equality axioms between terms,
 the notion of subterms is obscure.
We substitute this concept with the novel notion of places.
We assume to be given an infinite set of place symbols.

\begin{defi}\label{qbm79}
Suppose that a CCV $\lambda\mu$-term $M_0$ is given.
A term $M$ occurs
 at {\it place} $p$ if $p\mathbin{\hbox{\tt @}}M$ is derived
 by the following recursive process.
Let $p_0$ be an arbitrary place symbol.
We start with $p_0\mathbin{\hbox{\tt @}}M_0$, and apply the following
 operations recursively until we reach
 variables:
\begin{enumerate}[label=(\arabic*)]
\item
If either $p\mathbin{\hbox{\tt @}}(MN)$ or $p\mathbin{\hbox{\tt @}}
 (\uplet MxN)$, then $p\mathbin{\hbox{\tt @}}M$ and
 $q\mathbin{\hbox{\tt @}}N$ concurrently, where $q$ is a fresh place
 symbol.

\item
If either $p\mathbin{\hbox{\tt @}}(\lambda x.\,M)$ or
 $p\mathbin{\hbox{\tt @}}(\mu k.\,[l]M)$, then
 $q\mathbin{\hbox{\tt @}}M$, where $q$ is a fresh place symbol.
\end{enumerate}
We say that a place $q$ occurs in
 $M$ whenever $q\mathbin{\hbox{\tt @}}N$ is derived for some $N$
 from $p\mathbin{\hbox{\tt @}}M$ during the process
 (the case $p=q$ is inclusive).
We also say that $q\mathbin{\hbox{\tt @}}N$ occurs in $M$.
\end{defi}

Here we assume to read $[k](\uplet LxM)$ for $[k]\uplet LxM$.
For the associativity of the let-construct, either bracketing yields the
 same set of places up to the renaming of place symbols.
For instance, the following term has five places:
\[
\vbox{\offinterlineskip
 \halign{&\hfil $#$\hfil\cr
  &(\mu k.\,[l]&&(\lambda z.\,&&x)&&y)\uplet{}y{\,}&&x.\cr
	\noalign{\vskip2pt}
  \setplace{p_0} && \setplace{p_1} && \setplace{p_2}
   && \setplace{p_3} && \setplace{p_4}\cr
}}
\]
Namely, a place marks the location from which a term starts.
Note that one place may mark several terms.
For example, place $p_0$ marks both of the whole term and
 $\mu k.\,[l](\lambda z.\,x)y$.
Likewise, place $p_1$ marks $(\lambda z.\,x)y$ and
 $\lambda z.\,x$.
Each place marks all possible subterms starting at
 the position.

In the let-binding $\uplet MxN$, we regard that the scope $M$
 is viewable from the bound term $N$.
We extend the notion to future let-binding.
We consider $M$ to be visible from $N$ in
 term $\uplet Mx\mu k.\,\cdots[k]N$ though it is not currently
 the scope of $N$, as a
 one-step $\beta_\mu$-reduction yields
 $\mu k.\,\cdots [k]\uplet MxN$, in which 
 $M$ turns out to be the scope of $N$.
This idea naturally leads to the following definition.

\begin{defi}\label{fsq96}
Let $p$ be a place occurring in a given term $M_0$.
The {\it vision} $V(p)$ is
 the set of places in $M_0$ recursively defined as follows:

\begin{enumerate}[label=(\arabic*)]
\item
If $p\mathbin{\hbox{\tt @}}M$ marks the bound term
 of $p_1\mathbin{\hbox{\tt @}}\uplet LxM$,
 then $V(p)$ is defined by $\bigcup_q(\{q\}\cup V(q))$, where
 $q$ ranges over the set of all places 
 in $L$ ($p_1$ inclusive).

\item
If $p\mathbin{\hbox{\tt @}}M$ is preceded by a jumper, as in
 $p_1\mathbin{\hbox{\tt @}}\mu k.\cdots [k]M\cdots$,
 then $V(p)$ is defined by $V(p_1)$.
Intuitively, place $p$ is superposed over $p_1$ and
 the intermediate places between $\mu k$ and the jumper
 $[k]$ are invisible.
If $k$ is not bound, we deal with $V(p)$ by the following third rule
 (thus $V(p)=\emptyset$).

\item
For all other cases, we set $V(p)=\emptyset$.
\end{enumerate}
\end{defi}

\begin{rema}\label{agw33}\hfill
\begin{enumerate}[label=(\arabic*)]
\item
The places in $V(p)$ occur physically to the left of $p$.
Hence $V(p)$ is defined by induction from left to right.

\item
By definition, visions are transitive.
That is, if $r\in V(q)$ and $q\in V(p)$, then $r\in V(p)$.

\item
The definition of visions is irrelevant to the bracketing
 of let-constructs for associativity.
Namely, if $p\mathbin{\hbox{\tt @}}N$ occurs in $\uplet{\uplet LxM}y
 N$ with $y\not\in L$,
 then either bracketing yields the same set $V(p)$ up to the renaming
 of place symbols.

\item
In contrast, the interchange law of $\mu$ and let
 is annoying.
If we returned to the identification of
 $\uplet{(\mu k.\,J)}xM$ and $\mu k.\,(\uplet JxM)$, where $k\not\in M$
 as commented in Rem.~\ref{nrk88}, then
 we would have the following double vision problem.
Suppose $p$ is the place of $M$ in
 $p_1\mathbin{\hbox{\tt @}}\mu k.\,[l]\uplet LxM$, where $k\not\in M$.
If we understand it to be $p_1\mathbin{\hbox{\tt @}}
 \uplet{(\mu k.\,[l]L)}xM$, place $p_1$
 is visible from $p$.
On the other hand, if we regard the expression as
 $p_1\mathbin{\hbox{\tt @}}\mu k.\,[l](\uplet LxM)$, the
 vision of $p$ skips place $p_1$, jumping from $[l]$.
Namely, the vision is affected by bracketing.
\end{enumerate}
\end{rema}

Let us write $q\prec p$ if $q$ is immediately visible from $p$.
In other words, $q\in V(p)$ holds,
 while $q\in V(r)$ and $r\in V(p)$ hold for no $r$.

\begin{rema}\label{bug79}
The relation $q\prec p$ holds if and only if
 either of the following two cases happens.
(i)
 $p\mathbin{\hbox{\tt @}}N$ is the place of the bound term of
 $\uplet LxN$, and $q$ occurs in $L$.
Furthermore, if $q$ occurs in a let-expression
 $\uplet PyQ$ inside $L$, $q$ lies in the argument side, $Q$.
(ii)
 $p\mathbin{\hbox{\tt @}}N$ is the place preceded by a jumper, as
 in $p_1\mathbin{\hbox{\tt @}}
 (\mu l.\,\cdots [l]N\cdots)$ with $q\prec p_1$.
\end{rema}

\begin{defi}\label{sys59}
The {\it breadth} $|p|$ of a place $p$ is defined by induction on a physical
 location from left to right.
We define $|p|$ as the smallest natural number $n$ satisfying
 $|q|<n$ for all places $q\in V(p)$.
In particular, $|p|=0$ if $V(p)=\emptyset$.
\end{defi}

In other words, $|p|$ is the height of the tree of all sequences
 of places $q\prec q'\prec\cdots\prec p$ having $p$ as its root.

The following is an example of visions and breadths.
Let us consider
\[
\vbox{\offinterlineskip
 \halign{&\hfil $#$\hfil\cr
  &u&&v\uplet{}{u}{\,}&&(\mu l.[l]&&\uplet{v)}{v}{}&&\mu k.[k]&&
   (\lambda x.&&\mu m.[k]&&x).\cr
  \setplace{\vrule width0pt\kern-5ptp_0} && \setplace{p_1} && \setplace{p_2}
   && \setplace{p_3} && \setplace{p_4} && \setplace{p_5} && \setplace{p_6}
   && \setplace{p_7}\cr
}}
\]
Then $V(p_0)=V(p_1)=V(p_6)=\emptyset$ and $V(p_2)=V(p_3)=\{p_0,p_1\}$, while
$V(p_4)=V(p_5)=V(p_7)=\{p_0,p_1,p_2,p_3\}$.
Hence $|p_0|=|p_1|=|p_6|=0$ and $|p_2|=|p_3|=1$, while
$|p_4|=|p_5|=|p_7|=2$.
Let us suppose that a one-step $\beta_\mu$-reduction is applied to this
 term.
There are three possibilities, one for $\mu l$ and two for $\mu k$
 by the choices of bracketing.
One of the two for $\mu k$ inhales
 $\uplet{(\mu l.[l]v)}v{\square}$, yielding
\[
\vbox{\offinterlineskip
 \halign{&\hfil $#$\hfil\cr
   &u&&v\uplet{}u{}&&\mu k.[k]&&(\mu l.[l]&&v)\uplet{}v{}&&(\lambda x.&&\mu m.
   [k]&&(\mu l.[l]&&v)\uplet{}v{}&&x)\cr
  \setplace{\vrule width0pt\kern-5pt\bar p_0} && \setplace{\bar p_1}
   && \setplace{\bar p_4} && \setplace{\bar p'_2} && \setplace{\bar p'_3}
   && \setplace{\bar p_5} && \setplace{\bar p_6} && \setplace{\bar p''_2}
   && \setplace{\bar p''_3} && \setplace{\bar p_7}\cr
}}
\]
 where we write bars over places for distinction.
 The numbers associated with the places are given
 so that the correspondences are clear.
For example, $\bar p'_3$ and $\bar p''_3$ marking $v$ in the copied terms
 correspond to $p_3$ in front of $v$ before the reduction.
For instance,
 we have $V(\bar p''_3)=\{\bar p_0,\bar p_1\}$ and
$V(\bar p_5)=\{\bar p_0,\bar p_1,\bar p'_2,\bar p'_3\}$, while
$V(\bar p_7)=\{\bar p_0,\bar p_1,\bar p''_2,\bar p''_3\}$.
Note that the visions are essentially unaffected
 by the reduction.
This is because the reduction simply shifts the copies of
 $\mu l.[l]v$ to the points, where jumpers $[k]$ point to the
 original location skipping the intermediate parts.
The only exception is the place in front of $\mu k$.
We have $V(\bar p_4)=\{\bar p_0,\bar p_1\}$, thus $|\bar p_4|=1$, which
 decrease from $|p_4|=2$.

We consider the general case of a one-step $\beta_\mu$ reduction.
Let us suppose that the places are specified
 as follows.
 \[
\vbox{\offinterlineskip
 \halign{&\hfil $#$\hfil\cr
  &\uplet Mx{\,}&&\mu k.\,\cdots [k]&&N\cdots&
   \qquad\longrightarrow\qquad&&\mu k.\,\cdots [k]&&\uplet Mx{\,}&&N\cdots\cr
	\noalign{\vskip2pt}
  \setplace{s} && \setplace{r_0} && \setplace{t}
   &&&\setplace{\bar r_0} && \setplace{\bar s^{(i)}} && \setplace{\bar t}\cr
}}
\]
 where the displayed jumper denotes the $i$-th occurrence of $[k]$
 under an appropriate enumeration, $i=1,2,\ldots,n$.
For each place after the reduction, there is a unique
 corresponding place $p$ before it.
There are $n$ places $\bar p^{(i)}$ ($i=
 1,2,\ldots,n$), one in each copy of $M$, which correspond to $p$ in $M$.
Namely, we have an $n$-to-one correspondence if we look from
 the term after the reduction.
For the places $\bar p$ occurring elsewhere, there is
 a one-to-one correspondence to the places $p$ before the reduction.

We let the notation $C[M]$ mean a term wherein
 $p\mathbin{\hbox{\tt @}}M$ occurs for some $p$.

\begin{lem}\label{vcn44}
Let us consider the $\beta_\mu$ reduction $C[\uplet Mx{\mu k.\,J}]
 \rightarrow C[\mu k.\,J']$, where $J'=J\{[k]\square\mapsto [k]\uplet
 Mx\square\}$.
We set the place symbols as above.

\begin{enumerate}[label=(\arabic*)]
\item
The strict inequality $|\bar r_0|<|r_0|$ holds.

\item
$|\bar p|\leq |p|$ holds for every place $\bar p$
 occurring in $C[\mu k.\,J']$, where $p$ denotes
 the place in $C[\uplet Mx{\mu k.\,J}]$ associated
 with $\bar p$ by
 the ($n$-to-one and one-to-one) correspondence explained immediately
 above
(read $\bar p=\bar p^{(i)}$ for some $i$
 if it occurs in a copy of $M$).
\end{enumerate}
\end{lem}

\begin{proof}
  \begin{enumerate}[label=(\arabic*)]
\item is evident since $s\in V(r_0)$.
We note that $V(s)$ and $V(\bar r_0)$ has a one-to-one correspondence,
 and thus $|s|=|\bar r_0|$.

 \item 
Recall that the breadth equals the height of the tree
 of places ordered by $\prec$.
It suffices
 to show that $\bar q\prec \bar p$ implies
 $q\prec p$, excluding the case $\bar p=\bar r_0$.\quad
(i) The case that $\bar p$ occurs in one
 of the copies of $M$.
If $\bar p=\bar s^{(i)}$ in front of a copy of
 $M$, then $\bar q\prec \bar r_0$
 holds, as $\bar p$ is superposed over $\bar r_0$.
By the one-to-one correspondence between $V(\bar r_0)$ and $V(s)$, we
 have $q\prec s$.
Otherwise, $\bar p$ occurs strictly inside one of the copies
 of $M$ (namely, it is $\bar p^{(i)}$ for some $i$).
Then, $\bar q$ stays inside the same copy of $M$, or to
 the left of $\bar r_0$ when $\bar p$ is preceded by a jumper.
Namely, $\bar q$ never lies in the dotted part between $\bar r_0$
 and $\bar s^{(i)}$.
Hence $q\prec p$ holds from the beginning.\quad
(ii)
The case that $\bar p$ occurs in none of the copies of $M$.
If $\bar p=\bar t$, then  $\bar q$ occurs in $M$.
Now $q\prec r_0$ holds.
Therefore, $q\prec t$ as $t$ superposes over $r_0$.
If $\bar p\neq \bar t$, then $\bar q$ does not lie in $M$.
This follows from Rem.~\ref{bug79}, since
 $M$ is not a let-argument
 and none of the jumpers to the right of $\bar t$
 points into $M$, as the scope of $\mu$-binding in $M$
 must stay inside $M$.
As both $\bar q$ and $\bar p$ stay outside $M$,
 places $q$ and $p$ are not affected by the reduction.
Hence $q\prec p$ from the beginning. \qedhere
\end{enumerate}
\end{proof}

Next, we consider the $\beta_{\it jmp}$-reduction $C[[l]\mu k.\,J]\rightarrow
 C[J\{l/k\}]$.
Each place $\bar p$ in $C[J\{l/k\}]$ occurs either in $C$ or in $J\{l/k\}$.
Hence, we can naturally associate a place $p$ in $C[[l]\mu k.\,J]$
 that locates in $C$ or in $J$.

\begin{lem}\label{hrh81}
Let us consider the $\beta_{\it jmp}$-reduction $C[[l]\mu k.\,J]\rightarrow
 C[J\{l/k\}]$.
With each place $\bar p$ in $C[J\{l/k\}]$ is associated $p$ in $C[
 [l]\mu k.\,J]$, as explained immediately above.
Then $|\bar p|=|p|$ holds.
\end{lem}

\begin{proof}
A crucial case is when the place $p$ is preceded by a jumper $[k]$ in $J$.
Then, $\bar p$ is preceded by $[l]$.
If $l$ is bound and $\bar r_0$ is the place of $\mu l\cdots$ occurring
 in the context $C$, then $|\bar p|=|\bar r_0|$.
On the other hand, if we let $r_1$ denote the place of $\mu k.\,J$, we have
 $|p|=|r_1|=|r_0|$.
Evidently, $|\bar r_0|=|r_0|$,
 so $|\bar p|=|p|$.
If $l$ is not bound, we have $|\bar p|=0=|p|$.
\end{proof}

Next, we consider the $\eta_{\mu}$-reduction $C[\mu k.\,[k]M]\rightarrow
 C[M]$.
Each place $\bar p$ in $C[M]$ occurs either in $C$ or in $M$.
There is a corresponding place $p$ in $C[\mu k.\,[k]M]$
 located in $C$ or in $M$.

\begin{lem}\label{wnb98}
Let us take $\eta_{\mu}$-reduction $C[\mu k.\,[k]M]\rightarrow
 C[M]$.
As explained above, a place $p$ in $C[
 \mu k.\,[k]M]$ is associated with each place $\bar p$ in $C[M]$
Then $|\bar p|=|p|$ holds.
\end{lem}

\begin{proof}
Let $\bar r_1$ denote the place of $M$ after reduction.
Moreover, let $r_0$ denote the place of $\mu k.\,[k]M$.
Then, $|\bar r_1|=|r_0|=|r_1|$.
For other places, the lemma is immediate.
\end{proof}

We call $p$ a {\it $\mu$-place}
 if $p\mathbin{\hbox{\tt @}}\mu k.\,J$ happens.

\begin{defi}\label{nib98}
The {\it sight} of a term $M$ is the natural sum of $\omega^{|p|}$,
 where $p$ ranges over all $\mu$-places occurring in $M$.
\end{defi}

Written in a Cantor normal form, the sight of $M$ is
 equal to $\omega^{n_1}k_1+\omega^{n_2}k_2+\cdots +\omega^{n_s}k_s$
 for integers $k_i>0$ and $n_1>n_2>\cdots >n_s\geq 0$, where
 $k_i$ is the number of $\mu$-places $p$ satisfying $|p|=n_i$.

\begin{prop}\label{rpb09}
If $M_0\rightarrow M_1$ by an application of rule $\beta_\mu$,
 $\beta_{\it jmp}$, or $\eta_\mu$,
 the sight of $M_1$ is strictly less than the sight of $M_0$.
\end{prop}

\begin{proof}
For the $\beta_\mu$-reduction, we follow the notation in Lem.~\ref{vcn44}.
By (2) of the lemma, thhe breadth of each place never increases.
Each $\mu$-place $p$ in $M$ has $n$ copies after reduction.
However, we have $|p|<|r_0|$ since $p\in V(r_0)$.
To summarize, the places of breadth less than $|r_0|$ may be copied,
 while the places of breadth greater than or equal to $|r_0|$ are
 never copied.
Moreover, the breadth of $r_0$ itself decreases by (1) of the same lemma.
Hence the sight diminishes.

For the $\beta_{\it jmp}$-reduction, we use symbols in the proof of
 Lem.~\ref{hrh81}.
The breadth of $p$ never changes unless $p=r_1$.
Moreover, the place $r_1$ just vanishes.
For the $\eta_\mu$-reduction, we use Lem.~\ref{wnb98}.
The place $r_0$ vanishes.
\end{proof}

\begin{cor}\label{pxj93}
All $\beta_\mu\beta_{\it jmp}\eta_\mu$-reduction sequences are finite.
\end{cor}

\begin{proof}
Transfinite induction up to $\omega^\omega$ by Prop.~\ref{rpb09}.
\end{proof}

\begin{rema}\label{orv02}
To understand the necessity of regarding the exchange
 of the $\mu$-operator and
 let-operator as reduction, we consider
 the following example:
 \[
\uplet Kx{\uplet Ly{\uplet{\mu k.\,[m]M}z{\mu l.\,[k]N}}}.
\]
We assume $k$ does not occur elsewhere.
By applying a $\beta_\mu$-reduction to $\mu k$, we obtain
$\mu k.\,[m]\uplet Mz{\mu l.\,[k](\uplet Kx{\uplet LyN})}$.
Suppose $l\not\in N$.
If we can exchange $\mu$ and let, the term is equal to
\[
 \mu k.\,[m]\uplet Mz{\uplet{(\mu l.\,[k]K)}x{\uplet LyN}}.
\]
Now $M$ turns out to be visible from $L$,
 while it was previously out of vision,
 as $M$ occurred to the right of $L$.
Hence the vision of $L$ is widened by the reduction.
Furthermore, $M$ becomes visible from $N$, while
 it was previously skipped by jumper $[k]$.
These phenomena make the proof of Lem.~\ref{vcn44} fail.

This situation is troublesome since the problem does not arise
 if $l\in N$.
In this case, we cannot narrow the scope of $\mu k$.
The behavior is influenced by whether $l\in N$ or not.
We would need an argument sensitive to
 the occurrences of variables.
\end{rema}

\subsection{Characterization of strong normalizability}\label{eqw96}

We characterize strong normalizable terms in the CCV $\lambda\mu$-calculus
 by the union-intersection type discipline and by the CPS
 translation.
Strong normalizability is not closed under $\beta\eta$-convertibility.
So we must be sensitive to the choice of the CPS.
We assume the CPS defined via colon translation.
At the end of this section,
 we sketch how to generalize
 the results to the standard translation

\begin{prop}\label{ntw00}
Let $M$ be a term of the CCV $\lambda\mu$-calculus.
If $M$ is strongly normalizable, $[\![M]\!]$ is strongly normalizable.
\end{prop}

\begin{proof}
Toward contradiction, we assume that $M$ is strongly normalizable,
 while $[\![M]\!]$ admits an infinite reduction sequence.
By Lem.~\ref{jzz08},
 we have an infinite sequence of practical reductions
 from $[\![M]\!]^{-1}$ with no use of the rule {\it exch}.
By Cor.~\ref{pxj93}, we cannot continue
 $\beta_\mu$, $\beta_{\it jmp}$, and $\eta_\mu$ ceaselessly.
Hence there must be an infinite number
 of reductions other than these four rules.
By Lem.~\ref{qck48} and \ref{bdf72},
 an infinite reduction sequence from $M^\dagger$ is thus induced.
This contradicts the strong normalizability of $M$ as $M\mor*M^\dagger$.
\end{proof}

\begin{lem}\label{yiq64}
In the target calculus, if $T$ is strongly normalizable,
 then there is a derivation tree of $\Pi,\,\Theta\;\vdash_s\;
 T\mathbin:\tau$ for some $\Pi,\Theta$, and $\tau$.
Similar results hold for other sorts.
\end{lem}

\begin{proof}
This lemma is essentially the same as Cor.~3.4.4 of
 \cite[p.~159]{bkw61}.
We note that, in the ordinary lambda calculus, strong normalizability
 with respect to $\beta\eta$ obviously implies strong normalizability
 with respect to $\beta$.
As it can be easily checked, eacn
 term in $\beta$-normal form admits a type.
Then, by induction on the length of the longest reduction sequences,
 we can verify the lemma.
\end{proof}

\begin{prop}\label{xub13}
Let $M$ be a CCV $\lambda\mu$-term.
If $M$ is strongly normalizable, there is a derivation tree of
 a typing judgment $\Gamma\;\vdash\;M\mathbin:T\;|\;\Delta$
 for some $\Gamma,\Delta$, and $T$.
\end{prop}

\begin{proof}
By Prop.~\ref{ntw00}, $[\![M]\!]$ is strongly normalizable.
Hence, by Lem.~\ref{yiq64}, we obtain a derivation
 tree $\Pi,\,\Theta\;\vdash_s\;[\![M]\!]\mathbin:\tau$.
By Thm.~\ref{cnp96}, we have $\Gamma\;\vdash\;M\mathbin:T
 \;|\;\Delta$.
\end{proof}

\noindent
The main contribution of this section is the verification of
 the inverse of Prop.~\ref{ntw00} and \ref{xub13}.
To this end, we elaborate a syntactic translation that
 satisfies a kind of soundness with respect to reduction.
In the literature, we can find several proofs of strong
 normalizability by syntactic translations for call-by-value
 calculi with control operators \cite{vgu27}\cite{pug23}\cite{yis80}.
Though all use CPS translations, they have to manage the phenomenon
 that continuations may be discarded.
All of these works add some twists to the translations to avoid this problem.
Nakazawa preprocesses certain terms in the source calculus before
 the translation, carefully specifying harmful parts in the translation.
Ikeda and Nakazawa modify the CPS translation by adjoining
 extra terms they call garbage.
Kameyama and Asai use a two-level $\lambda$-calculus as
 the target of the translation to isolate the segment
 that does not discard continuations.
See Rem.~\ref{shw94} for more information.

Our translation adds garbage, inspired by \cite{pug23}.
In order to deal with the complication causeed by
 associative let-binding, 
 we add a new infix binary operator
 $(\hbox{-})\cdot(\hbox{-})$ to the ordinary lambda calculus.
We assume that the operator is syntactically associative:
\[
(L\cdot M)\cdot N\ =\ L\cdot (M\cdot N).
\]
We write $L\cdot M\cdot N$ for either of the two bracketings.
We consider the standard $\beta\eta$-reduction.
As a new reduction rule involving the dot operator,
 we add
 \[
M\cdot N\quad\rightarrow\quad N.
\]
We define a new type of colon translation.
The target calculus is the
 lambda calculus augmented by the binary dot operator.

\begin{defi}\label{cfa65}
We associate a fresh
 variable $\tilde k$
 with each of continuation variables $k$ in a one-to-one manner.
We define $\tilde K$ as follows:
\[
\tilde K = \begin{cases}%
  \smash{\tilde k} &\text{if } K=k \\
  Q & \text{if } K=\lambda x.\,Q\end{cases}.
\]
Each occurrence of variable $x$ becomes free in the second case.
We emphasize that $\tilde k$ is a variable independent from $k$.
So the substitution $k\mapsto K$ does not
 automatically substitute $\tilde k$ with $\tilde K$.
\end{defi}

\begin{defi}\label{sea85}
Let $M$ be a term in the CCV $\lambda\mu$-calculus, and let
 $K$ be a lambda term.
The lambda terms $\lg M\rg[K],\lg J\rg$, and $V^*$
 are simultaneously defined as in the following table:
\[
\halign{\kern5em $#$\hfil &${}\ =\ #$\hfil\cr
 \lg V\rg[K] & KV^*\cr
 \lg \uplet LxM\rg[K] & \lg L\rg[K]\cdot \lg M\rg[\lambda x.\,\tilde K\cdot
  \lg L\rg[K]]\cr
 \lg V_1V_2\rg[K] & V_1^*K\tilde KV_2^*\cr
 \lg VN\rg[K] & \tilde K\cdot\lg \uplet{Vz}zN\rg[K]\cr
 \lg NM\rg[K] & \tilde K\cdot\lg \uplet{zM}zN\rg[K]\cr
 \lg \mu k.\,J\rg[K] & \tilde K\cdot\lg J\rg\{K/k,\,\tilde K/\tilde k\}\cr
	\noalign{\vskip1ex}
 \lg [k]M\rg & \tilde k\cdot \lg M\rg[k]\cr
 \lg \uplet JxM\rg & \lg J\rg\cdot \lg M\rg[\lambda x.\,\lg J\rg]\cr
	\noalign{\vskip2ex}
 x^* & x\cr
 (\lambda x.\,M)^* & \lambda k\tilde kx.\,\lg M\rg[k]\cdot
  ((\lambda x.\,\tilde k\cdot \lg M\rg[k])x)\cr
}
\]
 where $V$ denotes a value and $N$ a non-value.
$L$ and $M$ are understood to be arbitrary terms, while
 $z$ is a fresh variable.
We assume the infix dot operator has higher precedence than
 lambda binding.
Though the definition is not a simple induction on construction,
 well-definedness is easy.

The obtained lambda terms do not follow the rules of sorts
 given in \S1.1.
We may renew the definition to accommodate
 the new colon translation.
The sorts, however, scarcely play a role
 hereafter, as inverse translation
 will not involved.
We use symbols such as $K$ and $Q$ only to indicate
 correspondence to the original target calculus.
\end{defi}

\begin{rema}\label{ran04}
As easily seen from Def.~\ref{sea85}, the associated
 $\tilde K$ actually occurs in $\lg M\rg[K]$, although $K$ may vanish.
\end{rema}

\begin{rema}\label{shw94}
We compare Def.~\ref{sea85} with the colon translation given in
 Preliminaries.
A crucial difference is that $K$ is actually substituted
 in the definition of the translation of $\mu k.\,J$.
This type of translations is found in
 the verification of strong normalizability
 \cite{vgu27}\cite{pug23}.
In fact, the first attempt by Parigot for the call-by-name
 $\lambda\mu$-calculus already used a translation where the continuation
 was actually substituted \cite{chq80}
 (unfortunately, the proof has a flaw; see \cite{ukg60}).
If $k$ does not occur in $J$, the substituted $K$ vanishes.
This is why we add the prefix $\tilde K$.
That is, we record the history of the
 continuations $K$, that may be deleted.
For a technical reason, the order of a value and a continuation
 is reversed in the translation of $V_1V_2$, and 
 the garbage $\tilde K$ is added.
The complication of $(\lambda x.\,M)^*$ is nothing more than
 for proof to work out.
\end{rema}

\begin{lem}\label{jcl39}
Let $M_1$ and $M_2$ be CCV $\lambda\mu$-terms.
If $M_1=M_2$ holds, then
 $\lg M_1\rg[K]=\lg M_2\rg[K]$ holds for every $K$.
Likewise, if $J_1=J_2$, then $\lg J_1\rg=\lg J_2\rg$ holds.
\end{lem}

\begin{proof}
First, both $\lg\uplet Lx{(\uplet MyN)}\rg[K]$ and
 $\lg\uplet{(\uplet LxM)}yN\rg[K]$ are equal to
 $Q'\cdot Q''\cdot\lg N\rg[\lambda y.\,\tilde K\cdot Q'
 \cdot Q'']$ where we set $Q'=\lg L\rg[K]$ and $Q''=
 \lg M\rg[\lambda x.\,\tilde K\cdot Q']$.
Second, both $\lg[k](\uplet LxM)\rg$ and $\lg \uplet{([k]L)}xM\rg$
 are equal to $\tilde k\cdot Q'\cdot \lg M\rg[\lambda x.\,\tilde k
 \cdot Q']$ for $Q'=\lg L\rg[k]$.
We comment that, in these two cases, the associativity of
 the infix dot operator is indispensable.
Since the definition in \ref{sea85} is compositional, the lemma follows.
\end{proof}

\begin{rema}\label{zny69}
For a fresh variable $k$, the equality $(\lg M\rg[k])\{K/k,\,
\tilde K/\tilde k\}=\lg M\rg[K]$ holds as naturally supposed.
We note that the definition of $\lg M\rg[k]$ contains $\tilde k$ implicitly.
Therefore, substituting only $k$ with $K$ does not suffice.
\end{rema}

\begin{lem}\label{lud98}
Let us put $\lg M\rg[K]=Q$ where $x\not\in K$,
 and let us consider two substitutions,
 $\theta_0=\{k\mapsto K,\>\tilde k\mapsto \tilde K\}$
 and $\theta=\{k\mapsto \lambda x.\,\tilde K\cdot Q,\;
 \tilde k\mapsto \tilde K\cdot Q\}$.
Then $\lg J\rg\theta=\lg J\{[k]\square\mapsto [k]\uplet Mx\square\}\rg\theta_0$
 holds.
\end{lem}

\begin{proof}
The proof is by induction on the construction of terms and jumps.
We simultaneously verify
 $V^*\theta=(V
 \{[k]\square\mapsto [k]\uplet Mx\square\})^*\theta_0$ and
 $\lg L\rg\theta[K']=\lg L
 \{[k]\square\mapsto [k]\uplet Mx\square\}\rg\theta_0[K']$.
Here the tricky notation $\lg L\rg\theta[K']$ means
 $\lg L\rg[K']\theta$ under the condition that $k,\tilde k\not\in K'$.
During the induction process, however, the variables $k$ and $\tilde k$ may occur
 in the position of $K'$.
In this case, $\lg L\rg[K']\theta$ should be manipulated
 as $\lg L\rg\theta[K'\theta]$.
An essential case is $J=[k]L$.
The left hand side is
 $\lg J\rg\theta=\tilde K\cdot Q\cdot \lg L\rg\theta[\lambda x.\,\tilde K
 \cdot Q]$ by Rem.~\ref{zny69}, while the right hand side equals
 $\tilde K\cdot Q\cdot \lg L\{[k]\square\mapsto[k]\uplet Mx\square\}\rg
 \theta_0[\lambda x.\,\tilde K\cdot Q]$.
Apply the induction hypothesis to~$L$.
\end{proof}

\begin{lem}\label{gph16}
Suppose $M_0\rightarrow M_1$, where the whole $M_0$ is a redex
 that is contracted.
Then $\lg M_0\rg[K]\mor{+}\lg M_1\rg[K]$ by one or more steps
 of $\beta\eta$ reduction (read $\lg J_0\rg\mor+\lg J_1\rg$
 for the case of $\beta_{\it jmp}$).
Moreover, $(\lambda x.\,Vx)^*\mor+V^*$ holds.
\end{lem}

\begin{proof}
First, consider the rule {\it exch}.
We have $\lg \uplet{(\mu k.\,J)}xM\rg[K]=\tilde K\cdot Q'\cdot
 \lg M\rg[\lambda x.\,\tilde K\cdot\tilde K\cdot Q']$, where
 $Q'=\lg J\rg\{K/k,\,\tilde K/\tilde k\}$.
On the other hand, $\lg \mu k.\,(\uplet JxM)\rg[K]$ is equal to
 $\tilde K\cdot Q'\cdot \lg M\rg[\lambda x.\,Q']$.
Therefore, the elimination of the
 two occurrences of $\tilde K$ settles this case.
We comment that $\lambda x.\,\tilde K\cdot\tilde K\cdot Q'$ actually occurs,
 in view of Rem.~\ref{ran04}.
Hence a positive number of $\beta$ reductions is enforced by the
 elimination.

For rule $\smash{{\it ad}_1}$, we have
 $\lg NM\rg [K]=\tilde K\cdot \lg \uplet{zM}zN\rg[K]$, regardless
 of whether $M$ is a value or a non-value.
Hence, eliminating $\tilde K$ yields $\lg \uplet{zM}zN\rg[K]$.
Rule $\smash{{\it ad}_2}$ is similarly handled.
For rule $\smash{\beta_\lambda}$, we have
 $\lg (\lambda x.\,M)V\rg[K]=(\lambda k\tilde kx.\,(\lg M\rg[k]\cdot
 ((\lambda x.\,\tilde k\cdot \lg M\rg[k])x
 )))K\tilde KV^*$.
Now, applying $\beta$ reduction to $k,\tilde k$, and $x$ yields
 $\lg M\rg[K]\cdot (\lambda x.\,\tilde K\cdot \lg M\rg[K])V^*$, which
 is equal to $\lg \uplet MxV\rg[K]$.
This finishes the case of $\smash{\beta_\lambda}$.
Furthermore, from the last term, eliminating $\lg M\rg[K]$
 and $\tilde K$ and applying a $\beta$ reduction to $x$
 yield $\lg M\rg[K]\{V^*/x\}$.
It is safe to assume that $K$ and $\tilde K$ contain
 no $x$, by $\alpha$-conversion if needed.
Therefore the last terms equals $\lg M\{V/x\}\rg[K]$.
This completes the case of rule $\smash{\beta_{\it let}}$.

For rule $\smash{\beta_\mu}$, we have $\lg \uplet Mx{\mu k.\,J}\rg[K]
 =Q\cdot \smash{\tilde K}\cdot Q\cdot\lg J\rg\theta$ where $Q$ and $\theta$
 are given in Lem.~\ref{lud98}.
Now the elimination of the two $Q$'s gives $\tilde K\cdot \lg J\rg\theta$.
It is equal to $\lg \mu k.\,J\{[k]\square\mapsto[k]\uplet Mx\square\}\rg[K]$
 by the same lemma, finishing this case.
We comment that $Q=\lg M\rg[K]$ is annihilated exactly
 at the moment of the dispatch of continuation
 $\uplet Mx\square$.
This ensures reductions antecedently done
 inside $M$ are not ignored even if $k\not\in J$.
For rule $\smash{\beta_{\it jmp}}$, we have $\lg [l]\mu k.\,J\rg
 =\tilde l\cdot \tilde l\cdot
 \lg J\rg\{l/k,\tilde l/\tilde k\}$.
Eliminating two $\tilde l$'s gives $\lg J\rg\{l/k,
 \tilde l/\tilde k\}=\lg J\{l/k\}\rg$.

For rule $\smash{\eta_{\it let}}$, we have $\lg \uplet xxM\rg[K]
 =Kx\cdot \lg M\rg[\lambda x.\,\tilde K\cdot Kx]$.
The elimination of $Kx$ and $\tilde K$ yields $\lg M\rg[\lambda x.\,Kx]$.
Now we apply the $\eta$ reduction to obtain $\lg M\rg[K]$.
For rule $\smash{\eta_\lambda}$, we must verify $(\lambda x.\,Vx)^*\mor+V^*$.
The left hand side equals $\lambda k\tilde kx.\,(V^*k\tilde kx)\cdot
 ((\lambda x.\,\tilde k\cdot V^*k\tilde kx)x)$.
Eliminating $V^*k\tilde kx$ and $\tilde k$ gives $\lambda k\tilde kx.\,
 (\lambda x.\,V^*k\tilde kx)x$.
Hence, one step of $\beta$ followed by three steps of $\eta$ yields $V^*$.
Thus $\lg \lambda x.\,Vx\rg[K]\mor+\lg V\rg[K]$ is also valid.
Finally, if $k\not\in M$, then $\lg \mu k.\,[k]M\rg[K]=\tilde K\cdot
 \tilde K\cdot \lg M\rg[K]$.
Eliminating two $\tilde K$'s gives $\lg M\rg[K]$, establishing the
 case of the $\eta_\mu$ rule.
\end{proof}

\noindent
Lemma~\ref{gph16} ensures the strict preservation of reductions
 in the case where the redex occurs naked at the topmost
 level.
In general, the redex $R$ may be encapsulated in a context as $C[R]$.
We introduce the notion of $E$-depth to handle this.

We recall the definition of places in Def.~\ref{qbm79}
 where the notion of an occurrence of
$q\mathbin{\hbox{\tt @}}N$ in $M$ was also defined.
\iffalse
Here we say that $q\mathbin{\hbox{\tt @}}N$
 occurs in $M$, specifying also the term $N$, whenever
 $q\mathbin{\hbox{\tt @}}N$ appears in
 the process from $p\mathbin{\hbox{\tt @}}M$.
\fi

In the following definition, we use
 the {\it evaluation contexts} $E$ defined by the following syntax:
 \[
E \mathrel{::=} \square\ \mid E[V\square] \mid E[\square M] \mid E[\uplet Mx\square].
\]
\begin{defi}\label{ets62}
The {\it E-depth} $d_M(q\mathbin{\hbox{\tt @}}L)$ is defined,
 when $q\mathbin{\hbox{\tt @}}L$ occurs in $M$.
As the base case, we set $d_L(q\mathbin{\hbox{\tt @}}L)=0$.
In general, the definition is provided by the following table, where
 $q\mathbin{\hbox{\tt @}}L$ is assumed to occur in $M$ or $V$.
\[
\halign{\kern5em $#$\hfil &${}\ =\ #$\hfil\cr
 d_{E[M]}(q\mathbin{\hbox{\tt @}}L) & d_M(q\mathbin{\hbox{\tt @}}L)\cr
 d_{E[VN]}(q\mathbin{\hbox{\tt @}}L) & 1+d_V(q\mathbin{\hbox{\tt @}}L)\cr
 d_{E[NM]}(q\mathbin{\hbox{\tt @}}L) & 1+d_M(q\mathbin{\hbox{\tt @}}L)\cr
 d_{E[\uplet MxN]}(q\mathbin{\hbox{\tt @}}L) & 1+d_M(q\mathbin{\hbox{\tt @}}L)\cr
 d_{\lambda x.\,M}(q\mathbin{\hbox{\tt @}}L) & 1+d_M(q\mathbin{\hbox{\tt @}}L)\cr
 d_{\mu k.\,[l]M}(q\mathbin{\hbox{\tt @}}L) & 1+d_M(q\mathbin{\hbox{\tt @}}L).\cr
}
\]
The second through the fourth of these equalities handle the
 case where $L$ occurs in the context $E$ if we split the term
 to the shape $E[M]$.
The number $d_M$ changes by bracketing of let-binding,
This does no harm, however,
 for we use the number only as the measure of complexity
 of terms to handle one-step reductions.
\end{defi}

We prove theorems by induction on the $E$-depth.
We first explore the base case, in which the redex $R$ has $E$-depth $0$.
Namely, the redex occurs in the form $E[R]$.
For rule $\beta_{\it jmp}$, we understand $R$ to
 be $\mu m.\,[l]\mu k.\,J$, including the $\mu$-operator
 preceding the redex $[l]\mu k.\,J$.
We must beware of rule $\eta_\lambda$, the redex of which is a value.

\begin{lem}\label{xap78}
Given an evaluation context $E$ and a term $K$ in the
 target calculus, there are $Q_E$ and $\mathcal{K}_E$ that
 satisfy the following:

\begin{enumerate}[label=(\arabic*)]
\item
If $N$ is a non-value, $\lg E[N]\rg[K]=Q_E\cdot \lg N\rg[\mathcal{K}_E]$.

\item
If $V$ is a value, $Q_E\cdot \lg V\rg[\mathcal{K}_E]\mor*\lg E[V]\rg[K]$.
\end{enumerate}

\noindent
Here $Q_E$ and $\mathcal{K}_E$ are terms in the extended language,
 although the former may be void.
If this is the case, we just ignore the preceding $Q_E$ and the following dot.
\end{lem}

\begin{proof}
  \begin{enumerate}[label=(\arabic*)]
    \item
The definition of $Q_E$ and $\mathcal{K}_E$ will be read off
 from the following equalities:
\[
\begin{alignedat}[b]{2}
  \lg E[NM]\rg[K] =&&~ Q_E\cdot \tilde{\mathcal{K}}_E\cdot\lg zM\rg[\mathcal{K}_E]\cdot
  \lg N\rg[\lambda z.\,\tilde{\mathcal{K}}_E&\cdot\lg zM\rg[\mathcal{K}_E]]\\
    \lg E[VN]\rg[K] =&&~ Q_E\cdot \tilde{\mathcal{K}}_E\cdot\lg Vz\rg[\mathcal{K}_E]\cdot
  \lg N\rg[\lambda z.\,\tilde{\mathcal{K}}_E&\cdot\lg Vz\rg[\mathcal{K}_E]]\\
    \lg E[\uplet MxN]\rg[K] =&&~ Q_E\cdot \lg M\rg[\mathcal{K}_E]\cdot
  \lg N\rg[\lambda x.\,\tilde{\mathcal{K}}_E&\cdot\lg M\rg[\mathcal{K}_E]]
\end{alignedat}
\]
Namely, the following inductive definition works.
If $E=\square$, we set $\mathcal{K}_\square=K$ and $Q_\square$ as void.
For inductive cases,
\begin{align*}
  Q_{E[\square M]} =&~ Q_E\cdot \tilde{\mathcal{K}}_E\cdot\lg zM\rg[\mathcal{K}_E]
  & \mathcal{K}_{E[\square M]}
  =&~ \lambda z.\,\tilde{\mathcal{K}}_E\cdot\lg zM\rg[\mathcal{K}_E] \\
    Q_{E[V\square]} =&~ Q_E\cdot \tilde{\mathcal{K}}_E\cdot\lg Vz\rg[\mathcal{K}_E]
  & \mathcal{K}_{E[V\square]}
  =&~ \lambda z.\,\tilde{\mathcal{K}}_E\cdot\lg Vz\rg[\mathcal{K}_E] \\
    Q_{E[\uplet Mx\square]} =&~ Q_E\cdot \lg M\rg[\mathcal{K}_E]
  & \mathcal{K}_{E[\uplet Mx\square]}
  =&~ \lambda x.\,\tilde{\mathcal{K}}_E\cdot\lg M\rg[\mathcal{K}_E]. 
\end{align*}
The first two in the left column depend on the choices of fresh variables $z$ in
 the construction of $\lg\cdot\rg$.

 \item
By induction on the construction of $E$.
In the case $E=\square$ and $E[\uplet Mx\square]$,
 both sides are equal.
In the case $E[\square M]$, by definition, $Q_{E[\square M]}\cdot
 \lg V\rg[\mathcal{K}_{E[\square M]}]$ reduces to $Q_E\cdot\lg
 \uplet{zM}zV\rg[\mathcal{K}_E]$ by dropping
 a single $\smash{\tilde{\mathcal{K}}_E}$.
By Lem.~\ref{gph16}, it reduces to $Q_E\cdot\lg VM\rg[\mathcal{K}_E]$, viz.,
 $\lg E[VM]\rg[K]$.
The case of $E[V'\square]$ is similar.\qedhere
\end{enumerate}
\end{proof}

\begin{lem}\label{fvk97}
Let $R\rightarrow S$ be one of the reduction rules.
Then, $\lg E[R]\rg[K]\mor+\lg E[S]\rg[K]$ with one or more steps
 of $\beta\eta$ reduction.
\end{lem}

\begin{proof}
Except for the $\eta_\lambda$-redex, $R$ is a non-value.
By Lem.~\ref{xap78}, (1), $\lg E[R]\rg[K]
 =Q_E\cdot \lg R\rg [\mathcal{K}_E]$, which contracts
 to $Q_E\cdot \lg S\rg[\mathcal{K}_E]$
 by one or more steps by Lem.~\ref{gph16}.
If $S$ is a non-value, we are done.
If $S$ is a value, we employ Lem.~\ref{xap78}, (2).
If $R$ equals $\lambda x.\,Vx$, which is a value, we need to take special care.
In the case of $E[\square N]$, the translation $\lg E[(\lambda x.\,Vx)N]\rg[K]$
 is as computed in the proof of Lem.~\ref{xap78}.
We apply $(\lambda x.\,Vx)^*\mor+V^*$, verified in Lem.~\ref{gph16},
 to the two occurrences of $\lg (\lambda x.\,Vx)z\rg[\mathcal{K}_E]
 =(\lambda x.\,Vx)^*\mathcal{K}_E\tilde{\mathcal{K}}_Ez$.
The other cases are similar.
\end{proof}

So the base case is done.
Now, by induction on $E$-depth, we can verify a central proposition
 of this subsection.

\begin{prop}\label{lfh56}
If $L\mor+M$ holds in the CCV $\lambda\mu$-calculus, then
 $\lg L\rg[K]\mor+\lg M\rg[K]$ holds with respect to $\beta\eta$ reduction
 for every $K$.
\end{prop}

\begin{proof}
It suffices to prove the case of the one-step reduction $L\rightarrow L_1$.
Let $R\rightarrow S$ be the instance of
 the reduction rule contracted by this step,
 and let $q\mathbin{\hbox{\tt @}}R$ be the place of the redex in $L$.
We show that if $d_L(q\mathbin{\hbox{\tt @}}R)\leq m$,
 then $L\rightarrow L_1$
 implies $\lg L\rg[K]
 \mor+ \lg L_1\rg[K]$, and simultaneously that if
 $d_V(q\mathbin{\hbox{\tt @}}R)\leq m$, then $V\rightarrow V_1$ implies $V^*
 \mor+V_1^*$ by induction on $m$.
We note that if $V\rightarrow V_1$ and if $V$ is a value, $V_1$ is also a value.
The base case $m=0$ is Lem.~\ref{fvk97}.
We verify the induction step.

\begin{enumerate}[label=(\roman*)]
\item First, we consider the case where $q\mathbin{\hbox{\tt @}}R$ occurs
 in the $V$ in $L=E[VM']$.
We split cases further according to whether $M'$ is a value.
If it is a value $W$, we have $\lg E[VW]\rg [K]=\tilde{\mathcal{K}}_E\cdot
 V^*\mathcal{K}_E\tilde{\mathcal{K}}_EW^*$.
Since $d_V(q\mathbin{\hbox{\tt @}}R)<d_{E[VW]}(q\mathbin{\hbox{\tt @}}R)$,
 we apply the induction hypothesis to $V^*$.
If $M'$ is a non-value $N$, the computation
 of $\lg E[VN]\rg[K]$ is displayed in Lem.~\ref{xap78}.
Note that $\lg Vz\rg[\mathcal{K}_E]=V^*\mathcal{K}_E\tilde{\mathcal{K}}_Ez$.
We apply the induction hypothesis to all occurrences of $V^*$.
Observe that at least one occurrence of $V^*$ exists.
\item
The case where $q\mathbin{\hbox{\tt @}}R$ occurs in the $M$ in $L=
 E[NM]$.
The computation of $\lg E[NM]\rg[K]$ is given in Lem.~\ref{xap78}.
We note $d_{zM}(q\mathbin{\hbox{\tt @}}R)=d_M(q\mathbin{\hbox{\tt @}}R)
 <d_{E[NM]}(q\mathbin{\hbox{\tt @}}R)$ since $z$ is a value while $N$ is
 a non-value.
Apply the induction hypothesis to all occurrences of $\lg zM\rg[\mathcal{K}_E]$.
\item
The case where $q\mathbin{\hbox{\tt @}}R$ occurs in the $M$ in
 $L=E[\uplet Mx{M'}]$.
The computation of $\lg E[\uplet Mx{M'}]\rg[K]$ is given in Lem.~\ref{xap78},
 if we read $N=M'$.
Since $d_M(q\mathbin{\hbox{\tt @}}R)<d_{E[\uplet Mx{M'}]}(
q\mathbin{\hbox{\tt @}}R)$, apply induction hypothesis to all
 occurrences of $\lg M\rg[\mathcal{K}_E]$.
\item
Case where $q\mathbin{\hbox{\tt @}}R$ occurs in $M$ of
 $L=E[\lambda x.\,M]$.
We have $\lg E[\lambda x.\,M]\rg[K]=Q_E\cdot \mathcal{K}_E[
 \lambda k\tilde kx.\,(\lg M\rg [k]\cdot ((\lambda x.\,\tilde k\cdot
 \lg M\rg[k])x))]$.
Apply induction hypothesis to the two occurrences of $\lg M\rg[k]$
 since $d_M(q\mathbin{\hbox{\tt @}}R)<d_{E[\lambda x.\,M]}(
 q\mathbin{\hbox{\tt @}}R)$.
We note that, taking $E$ to be void, this case essentially contains
 the proof of $V^*\mor+V_1^*$.
\item
Finally, we consider the case where $q\mathbin{\hbox{\tt @}}R$ occurs in the $M$ in
 $L=E[\mu k.\,[l]M]$.
We have $\lg E[\mu k.\,[l]M]\rg[K]=Q_E\cdot \tilde{\mathcal{K}}_E\cdot
 \tilde l\cdot \lg M\rg[l]\{\mathcal{K}_E/k,\,\tilde{\mathcal{K}}_E/\tilde k\}$.
Apply the induction hypothesis to $\lg M\rg[l]$.
\end{enumerate}
\end{proof}

We define the translation of the union-intersection types of
 CCV $\lambda\mu$-calculus into intersection types of the target calculus:
\begin{align*}
  &\alpha^* =~ \alpha, \quad (S \rightarrow T)^* = T^+ \rightarrow \dbot \rightarrow \neg S^*, \quad  (\textstyle{\bigcap}\, R)^* = \textstyle{\bigcap}\, R^* \\[.5em]
  &(\textstyle{\bigcup}\, S)^+ = \textstyle{\bigcap}\, \neg S^* \\[.5em]
  &[\![T]\!] = \neg T^+ 
\end{align*}
In the previous paper, we defined $(S\rightarrow T)^*$ as
 $S^*\rightarrow [\![T]\!]$.
The modification corresponds to the change of the colon translation.
Although the results of the type translation do
 not obey the rule of the
 target calculus in \S\ref{ehd63}, this does not matter in the following argument.
We simply ignore the distinction between $\sigma,\kappa$, and $\tau$.
Accordingly, there is no need to distinguish between the environments
 $\Pi,\Theta$.
We add the following inference rule:
\[
  \frac{\Pi \vdash_s Q_1\mathbin:\dbot\qquad \Pi \vdash_s Q_2\mathbin:\dbot}{\Pi
    \vdash_s Q_1\cdot Q_2\mathbin:\dbot}
\]

\begin{lem}\label{aei16}
If $\Gamma\;\vdash\;M\mathbin:T\;\mathbin|\;\Delta$ is derived
 in the CCV $\lambda\mu$-calculus, then $\Pi,\,k\mathbin:T^+\!,
 \,\tilde k\mathbin:\dbot
 \;\vdash_s\;\lg M\rg[k]\mathbin:\dbot$ is derived for some
 $\Pi$.
\end{lem}

\begin{proof}
This lemma was proved in our previous paper \cite{tsd43} for the original colon
 translation.
We follow the same line.
We need to pay attention, however, to the pieces added by the dot operator.
Let us consider the case
 $\Gamma\;\vdash\;\uplet JxM\mathbin:\dbot\;\mathbin|\;\Delta$
 inferred from $[\>\Gamma,\,
	\penalty-3000
x\mathbin:S_i\;\vdash\;J\mathbin:
 \dbot\;\mathbin|\;\Delta\>]_i$ and $\Gamma\;\vdash
 \;M\mathbin:\bigcup S_i\;|\;\Delta$.
We have $\lg \uplet JxM\rg[k]=\lg J\rg\cdot \lg M\rg[\lambda x.\,\lg J\rg]$.
We must take care of the free occurrences of $x$ that may appear in the 
 first $\lg J\rg$ in front of the dot.
Hence, we choose $i_0$ and add $x\mathbin:S_{i_0}^*$ to the typing environment.
We then do the same for $\uplet LxM$.
Next we consider the case $N_1N_2$, where $N_i$ are non-values.
We have $\lg N_1N_2\rg[k]=\tilde k\cdot \lg zN_2\rg[k]\cdot
 \lg N_1\rg[\lambda z.\,\tilde k\cdot \lg zN_2\rg[k]]$, where
 $\lg zN_2\rg[k]=\tilde k\cdot (zk\tilde kw)\cdot \lg N_2\rg
 [\lambda w.\,\tilde k\cdot (zk\tilde kw)]$.
The variable $z$ occurs freely in the left $\lg zN_2\rg[k]$, while
 the variable $w$ occurs freely in the left
 $zk\tilde kw$ contained in $\lg zN_2\rg[k]$.
We choose $i_0$ and add $z\mathbin:\bigcap_j(T^+\rightarrow\dbot\rightarrow
 \neg S_{i_0j}^*)$ to the type environment
Moreover, we choose
 $j_0(i)$ for each $i$ and also add
 $w\mathbin:\mathop{\smash{\bigcap_i}}\smash{S_{i\,j_0(i)}^*}$ to the type environment.
Then, from the induction hypotheses on $N_i$, we can infer the typing of
 $\lg N_1N_2\rg[k]$ under the augmented environment, in spite of
 the free occurrences of $z$
 and $w$.
The remaining cases are similar.
\end{proof}

\begin{prop}\label{yla37}
If a CCV $\lambda\mu$-term $M$ is typeable, then $M$ is strongly normalizable.
\end{prop}

\begin{proof}
If $M$ is typeable,
 $\lg M\rg[k]$ is typeable by Lem.~\ref{aei16}.
Then, as sketched in the Appendix, Cor.~\ref{rqh00},
 $\lg M\rg[k]$ is strongly normalizable
 in the lambda calculus with an additional rewriting rule
 for the dot operator.
Therefore $M$ is strongly normalizable by Prop.~\ref{lfh56}.
\end{proof}

\begin{thm}\label{rzq50}
A CCV $\lambda\mu$-term $M$ is strongly normalizable if and only
 if $M$ is typeable.
\end{thm}

\begin{proof}
A combination of Prop.~\ref{xub13} and \ref{yla37}.
\end{proof}

\begin{thm}\label{tml91}
A CCV $\lambda\mu$-term $M$ is strongly normalizable if and only
 if $[\![M]\!]$ is strongly normalizable.
\end{thm}

\begin{proof}
The only-if part is just Prop.~\ref{ntw00}.
The converse uses the characterization by types.
Provided that $[\![M]\!]$ is strongly normalizable,
 $M$ is typeable.
This is included in the proof of Prop.~\ref{xub13}.
Finally, $M$ is strongly normalizable by Prop.~\ref{yla37}.
\end{proof}

\begin{rema}\label{fpg06}
We call a CCV $\lambda\mu$-term strongly quasi-normalizable
 if all reduction sequences containing none of the $\eta$-type rules
 are finite.
Let us show that strong quasi-normalizability implies strong
 normalizability.
Suppose $M$ is strongly quasi-normalizable.
By inspection of the proof of Prop.~\ref{ntw00},
 we see that $[\![M]\!]$ is strongly $\beta$-normalizable.
Hence $[\![M]\!]$ is typeable with no use of $\omega$.
The rest of the proof goes as that of Thm.~\ref{tml91}.
Contrary to the case of weak normalizability \cite{tsd43},
 the inverse is trivial.
\end{rema}

\begin{rema}\label{zqb05}
Let us briefly discuss how to extend the main theorem
 to the standard CPS translation.
The translation is defined by the following:
\[
\halign{\kern5em $#$\hfil &${}\ \mathrel{:=}\ #$\hfil\cr
 [\![V]\!] & \lambda k.\,kV^*\cr 
 [\![MN]\!] & \lambda k.\,[\![M]\!](\lambda x.\,[\![N]\!](\lambda y.\,xyk))\cr
 [\![\uplet MxN]\!] & \lambda k.\,[\![N]\!](\lambda x.\,[\![M]\!]k)\cr
 [\![\mu k.\,J]\!] & \lambda k.\,[\![J]\!]\cr
 [\![[k]M]\!] & [\![M]\!]k\cr
 [\![\uplet JxN]\!] & [\![N]\!](\lambda x.\,[\![J]\!])\cr
 \noalign{\vskip1ex}
 x^* & x\cr
 (\lambda x.\,M)^* & \lambda x.\,[\![M]\!].\cr
}
\]
Theorem~\ref{tml91} remains valid for this definition.
This is verified along the following line.
For distinction, let $[\![M]\!]_c$ denote
 the CPS via the colon translation
 given in Preliminaries.
Let us observe that, if we modify the standard CPS translation
 by $[\![\mu k.\,J]\!]'=\lambda k.\,(\lambda k.\,[\![J]\!]')k$
 and by $(\lambda x.\,M)^*=\lambda xk.\,[\![M]\!]'k$, then
 we have $[\![M]\!]'\mor*[\![M]\!]_c$.
Here $[\![M]\!]'$ is $\eta$-equal to $[\![M]\!]$.
In the ordinary lambda calculus,
 strong normalizability is stable under $\eta$-equality
 (one way to verify this is to use the characterization by intersection types).
Hence, if $[\![M]\!]$ is strongly normalizable, so are $[\![M]\!]'$ and,
 in turn, $[\![M]\!]_c$.
Therefore $M$ is strongly normalizable by the theorem~\ref{tml91}.
For the only-if part, if $M$ is strongly normalizable, $M$ is typeable
 by Prop.~\ref{xub13}.
We can prove that the typeability of $M$ induces
 the typeability of $[\![M]\!]$
 (verified for $[\![M]\!]_c$ in \cite{tsd43}).
Hence $[\![M]\!]$ is strongly normalizable.
\end{rema}

\section{Conclusion}\label{bim82}

In the series of two papers, we presented call-by-value lambda
 calculi with control operators.
They are complete with respect to the standard CPS semantics.
The key idea is to introduce equality axioms between terms, departing
 from the convention that terms are freely generated by grammars.

We demonstrated the aptitude of the calculi
 through several mathematical properties.
In this second paper, we gave the characterization of the strong
 termination property of the CCV $\lambda\mu$-calculus.
We verified the following two results:
\begin{enumerate}[label=(\arabic*)]
  \item
$M$ is strongly normalizing iff its CPS translation $[\![M]\!]$ is
 strongly normalizing.
 \item
$M$ is strongly normalizing iff $M$ is typeable (with
 use of empty intersection or empty union nowhere).
\end{enumerate}
We mention future problems
 that are not tackled in our series of papers.
We adopted the reduction rule {\it exch} for the characterization
 of strong termination.
It remains open if we assume the equality rule \ref{nrk88}
 exchanging $\mu$ and let in place.
Second, we plan to extend the results to a
 $\lambda\mu$-calculus having delimited control operators.
As a matter of fact, this work is a precursory extract
 from our attempt to develop complete
 calculi with delimited control operators.

In the literature, we can find the characterization of
 strong normalizability for various systems of
 the $\lambda\mu$-calculus\footnote{%
We thank an anonymous referee who informed
 us of the related works.}.
Van~Bakel, Barbanera, and de'Liguoro
 considered a call-by-name $\lambda\mu$-calculus using
 special forms of intersection and product types \cite{efa77}.
Tsukada and Nakazawa introduced a polarized variation of
$\bar\lambda\mu\tilde\mu$-calculus and gave a
 characterization using union and intersection types \cite{rdt15}.
In particular, the latter may have a close connection to our results,
 although the details remain to be investigated in the future.

\appendix

\section{Strong normalizability of the extended lambda calculus}

We give a sketch of the strong normalizability result needed
 in the proof of Prop.~\ref{yla37}.
We extend the ordinary lambda calculus by a binary dot operator
 $M\cdot N$, which
is associative, i.e., $(L\cdot M)\cdot N=L\cdot (M\cdot N)$ holds.
We omit the brackets.
As a new reduction rule related to the dot operator, we add
\[
M\cdot N\quad\rightarrow\quad N
\]
in addition to the ordinary $\beta\eta$-reduction.

We introduce an intersection type system.
We do not need sorts.
So the strict types $\sigma$ and types $\tau=\underline\sigma$ are
 defined simply by
\begin{align*}
 \sigma \mathrel{::=}&~\alpha \mid \tau\rightarrow\sigma \\
  \tau \mathrel{::=} &~\textstyle{\bigcap}\,\sigma
\end{align*}
 where $\alpha$ ranges over atomic types and
 $\bigcap \sigma$ signifies a finite intersection
 $\sigma_1\cap\sigma_2\cap \cdots \cap\sigma_n$ ($n\geq 1$).
We emphasize that the nullary intersection $\omega$ is not
 considered here.
We assume that a special atomic type $\dbot$ is included.

We naturally define the subtype relation $\leq$.
As typing rules, we add
\[
\vbox{\offinterlineskip
 \halign{\strut $#$\hfil\cr
  \Gamma\ \vdash_s\ M\mathbin:\dbot\qquad \Gamma\ \vdash_s\ N\mathbin:\dbot\cr
  \noalign{\hrule}
  \Gamma\ \vdash_s\ M\cdot N\mathbin:\dbot\cr}}
\]
 to the standard rules, including the inheritance rule.
We verify the strong normalizability of this type system
 by the standard computability method \cite{bkw61}\cite{rne46}.

Strong normalizability satisfies the following three properties elucidated
 in \cite{rne46},
 even if we consider $\eta$-reduction and the new
 rule associated with the dot operator:\quad
(i) if $L_1,L_2,\ldots,L_n$ ($n\geq 0$) are strongly normalizable,
 $xL_1L_2\cdots L_n$ is strongly normalizable,\quad
(ii) if $Mx$ is strongly normalizable, $M$ is strongly normalizable.\quad
(iii) if $N$ and $M\{N/y\}L_1L_2\cdots L_n$
 ($n\geq 0$) are strongly normalizable, $(\lambda y.\,M)NL_1L_2\cdots
 L_n$ is strongly normalizable.

\begin{defi}\label{uep04}
We define a family of sets ${\it Comp}_\tau$ of terms by
 induction on the construction of type $\tau$.
\begin{align*}
 M\in {\it Comp}_\alpha \Longleftrightarrow&~ \hbox{$M$ is strongly normalizable}\\
 M\in {\it Comp}_{\tau\rightarrow\tau'}
  \Longleftrightarrow&~ \forall N\in{\it Comp}_\tau.\;MN\in{\it Comp}_{\tau'} \\
  M\in {\it Comp}_{\tau\cap\tau'} \Longleftrightarrow&~ M\in {\it Comp}_{\tau}\cap
  {\it Comp}_{\tau'}.
\end{align*}
It does no harm to consider non-typeable terms.
So we do not include the condition of types for simplicity.
\end{defi}

\begin{lem}\label{uoi75}
The following hold for each type $\tau$:
\begin{enumerate}[label=(\arabic*)]
\item $x\in {\it Comp}_\tau$.
\item If $M\in {\it Comp}_\tau$, then $M$ is strongly normalizable.
\end{enumerate}
\end{lem}

\begin{proof}
Simultaneous induction on construction of types $\tau$.
To let induction go through, we strengthen condition (1):
 $xM_1M_2\cdots M_n\in {\it Comp}_\tau$ whenever $M_1,M_2,\ldots,
 M_n$ are strongly normalizable.
To handle the case of $\tau\rightarrow\tau'$, we need property (i) given
 above.
For (2) of $\tau\rightarrow\tau'$, we need (ii).
We comment that the lemma fails if we allow the nullary intersection.
\end{proof}

\begin{lem}\label{mid32}
If $\tau\leq\tau'$ holds, ${\it Comp}_\tau\subseteq {\it Comp}_{\tau'}$ holds.
\qed
\end{lem}

\begin{lem}\label{brz11}
Suppose that $x_1\mathbin:\tau_1,\,x_2\mathbin:\tau_2,\,
 \ldots,\,x_n\mathbin:\tau_n\;\vdash_s\;M\mathbin:\sigma$ holds.
For every $n$-tuple of $P_i\in {\it Comp}_{\tau_i}$, we have
 $M\{P_1/x_1,P_2/x_2,\ldots,P_n/x_n\}\in {\it Comp}_\sigma$.
\end{lem}

\begin{proof}
Induction on the construction of
 the derivation trees of typing judgments.
We consider the rule deriving $M\cdot N\mathbin:\dbot$
 from $M\mathbin:\dbot$ and $N\mathbin:\dbot$.
The substitution $P_i/x_i$ plays no role in this case,
 so we omit it.
Since $\dbot$ is an atomic type, the induction hypotheses say
 $M$ and $N$ are strongly normalizable.
Then, $M\cdot N$ is obiously also strongly normalizable, that is,
 $M\cdot N\in {\it Comp}_{\sdbot}$.
To handle lambda-abstraction, we need property (iii) given above.
For the inheritance rule, we use Lem.~\ref{mid32}.
\end{proof}

\begin{cor}\label{rqh00}
If $\Gamma\vdash_sM\mathbin:\sigma$ is derivable, $M$ is strongly normalizable.
\end{cor}

\proof\hskip.5em
We take $x_i$ as $P_i$ in Lem.~\ref{brz11}.
Then, noting (1) of Lem.~\ref{uoi75},
 we have $M\in {\it Comp}_\sigma$.
Thus $M$ is strongly normalizable by (2) of Lem.~\ref{uoi75}.
\qed

%% in general the use of bibtex is encouraged

\end{document}